\documentclass[12pt]{article}

\usepackage{amssymb}
\usepackage{color}
\usepackage{epsfig,amssymb,amsfonts,amsmath,graphicx,dsfont,cite,xfrac}
\usepackage{authblk}
\usepackage{subcaption}
\definecolor{mygray}{gray}{0.5}

\usepackage{cite}
\usepackage[colorlinks=true,linkcolor=blue,citecolor=red]{hyperref}
\newcommand{\be}{\begin{equation}}
\newcommand{\ee}{\end{equation}}
\newcommand{\bea}{\begin{eqnarray}}
\newcommand{\eea}{\end{eqnarray}}

\title{Coherent and squeezed states: introductory review of basic notions, properties and generalizations}

\author[${}$]{Oscar Rosas-Ortiz}

\affil[${}$]{\footnotesize Physics Department, Cinvestav, AP 14-740, 07000 M\'exico City, Mexico}

\date{}
\begin{document}

\maketitle

\begin{abstract}
A short review of the main properties of coherent and squeezed states is given in introductory form. The efforts are addressed to clarify concepts and notions, including some passages of the history of science, with the aim of facilitating the subject for nonspecialists. In this sense, the present work is intended to be  complementary to other papers of the same nature and subject in current circulation. 
\end{abstract}

%%%%%%%%%%%%%%%%%%%%%%%%%%%%%%%%%%%%%%%%%%%%%%%

%--------------------------------------->

\section{Introduction}
\label{intro}

Optical coherence refers to the correlation between the fluctuations at different space--time points in a given electromagnetic field. The related phenomena are described in statistical form by necessity, and include interference as the simplest case in which correlations between light beams are revealed \cite{Man65}. Until the first half of the last century the classification of coherence was somehow based on the averaged intensity of field superpositions. Indeed, with the usual conditions of stationarity and ergodicity,  the familiar concept of coherence is associated to the sinusoidal modulation of the averaged intensity that arises when two fields are superposed. Such a modulation produces the extremal values $\langle I_{max} \rangle_{av}$ and $\langle I_{min} \rangle_{av}$, which are used to define the visibility of interference fringes 
\[
{\cal V} = \frac{\langle I_{max} \rangle_{av} - \langle I_{min} \rangle_{av}} {\langle I_{max} \rangle_{av} + \langle I_{min} \rangle_{av}}.
\]
The visibility is higher as larger is the difference $\langle I_{max} \rangle_{av} - \langle I_{min} \rangle_{av} \geq 0$. At the limit $\langle I_{min} \rangle_{av} \rightarrow 0$ we find ${\cal V} \rightarrow 1$. If no modulation is produced then $\langle I_{max} \rangle_{av} = \langle I_{min} \rangle_{av}$, and no fringes are observed (${\cal V} =0$). The fields producing no interference fringes are called {\em incoherent}. In turn, the highest order of coherence is traditionally assigned to the fields that produce fringes with maximum visibility. 

The Young's experiment is archetypal to introduce the above concepts. Let us write $2 \vert \Lambda_{(1,2)} \vert \cos \theta_{(1,2)}$ for the sinusoidal modulation of the averaged intensity at the detection screen. Then
\[
\langle  I_{min} \rangle_{av} = \langle I_1 \rangle_{av} + \langle I_2 \rangle_{av} - 2 \vert \Lambda_{(1,2)} \vert,  \qquad \langle I_{max} \rangle_{av} = \langle I_{min} \rangle_{av} + 4 \vert \Lambda_{(1,2)} \vert, 
\]
with $\langle I_1 \rangle_{av}$ and $\langle I_2 \rangle_{av}$ the average intensities that would be contributed by either pinhole in the absence of the other. In general $\langle I_1 \rangle_{av}$, $\langle I_2 \rangle_{av}$, and $\vert \Lambda_{(1,2)} \vert$ depend on the geometry of the experimental setup. Therefore we can write
\be
{\cal V}_Y = \left( \frac{2 \sqrt{ \langle I_1 \rangle_{av} \langle I_2 \rangle_{av}} }{\langle I_1 \rangle_{av} + \langle I_2 \rangle_{av}} \right) \lambda_{(1,2)}, \quad \mbox{with} \quad  \lambda_{(1,2)} = \frac{\vert \Lambda_{(1,2)} \vert}{\sqrt{ \langle I_1 \rangle_{av} \langle I_2 \rangle_{av}} }.
\label{young}
\ee
The expression $\lambda_{(1,2)}$ is the {\em correlation function} of the variable fields associated to the averages $\langle I_1 \rangle_{av}$ and $\langle I_2 \rangle_{av}$. If  $\vert \Lambda_{(1,2)} \vert =0$ then $\lambda_{(1,2)} =0$, and no fringes are observed (${\cal V}_Y =0$). The fields incident on the pinhole--screen are in this case {\em incoherent}. On the other hand, if the fields emerging from the pinholes have equal intensity, the visibility ${\cal V}_Y$ will be equal to 1 only when $\lambda_{(1,2)} =1$. The simplest form to obtain such a result is by considering the factorized form $\vert \Lambda_{(1,2)} \vert = \sqrt{ \langle I_1 \rangle_{av} \langle I_2 \rangle_{av} }$, together with $\langle I_1 \rangle_{av} = \langle I_2 \rangle_{av}$. 

The change of paradigm emerged in 1955, after the Brown and Twiss experiments oriented to measure correlations between quadratic forms of the field variables \cite{Bro56,Bro57,Bro58}. Unlike ordinary interferometer outcomes, the results of  Brown and Twiss demanded the average of  square intensities for their explanation. In other words, to embrace the new phenomenology, the concept of coherence as well as the {\em first-order} correlation function $\lambda_{(1,2)}$ needed a generalization. It was also clear that not all the fields described as ``coherent'' in the traditional approach would end up satisfying the new definitions of coherence. Thus, the Brown--Twiss results opened up the way to the quantitative investigation of higher-order forms of coherence \cite{Man65,Tro74,Man95,Bor99}, though most of the light studied at the time was mainly produced by thermal sources. The development of new sources of light (minimizing noise generation) and new detectors (strongly sensitive to individual quanta of light) represented the experimental sustenance for the study of such concepts. However, the latter implied that the formal structure of optical coherence should be constructed on the basis of two mutually opposed features of light. While interference has long been regarded as a signature of the wavelike nature of electromagnetic radiation (Maxwell theory), photodetection implies the annihilation of individual photons to release photoelectrons from a given material (Einstein description of the photoelectric effect). These contradictory aspects were reconciled by Glauber in 1963,  with his quantum approach to optical coherence, after considering expectation values of normal--ordered products of the (boson) creation and annihilation operators as quantum analogs of the classical correlation functions \cite{Gla63a,Gla63b,Gla63c}. The expectation value of a single normal--ordered product corresponds to the first-order correlation function, that of two products corresponds to the second-order correlation function, and so on. Basically, Glauber formulated the theory of quantum optical detection for which the Young and Brown--Twiss experiments correspond to the measurement of first- and second-order correlation functions, respectively \cite{Wal79}.

According to Glauber, most of the fields generated from ordinary sources lack second- and higher-order coherence, though they may be considered ``coherent'' in the traditional sense (i.e., they are only first-order coherent in Glauber's approach) \cite{Gla64}. Partial coherence means that there exist correlations up to a finite order only, and that the correlation functions of such order and all the lower orders are normalized. Full coherence implies `partial coherence' at all orders (the complete compilation of the Glauber contributions  to quantum theory of optical coherence can be found in \cite{Gla07}). In concordance with $\lambda_{(1,2)}$ in the Young's interferometer, the Glauber's approach adopts the factorization of the $n$-th-order  correlation function as a condition for coherence. Each of the factors corresponds to the probability of detecting a single photon at a given position at a given time. Factorization represents independence in the single-photon detection process. As recording a photon means its annihilation, the factors are indeed the squared norm of the vector state after the action of an ideal detector. At this stage, the brilliant contribution of Glauber was to notice that the simplest form of satisfying full coherence is by asking the quantum state of the field to be an eigenvector of the boson-annihilation operator with complex eigenvalue. Quite interestingly, this eigenvalue is a solution (written in complex form) of  the corresponding Maxwell equations. In this form, the Glauber's fully coherent states of the quantized electromagnetic radiation are directly associated to the conventional electromagnetic theory. A classical description is then feasible for such a special class of quantum states of light.

It is apparent that the Fock (or number) states $\vert n \rangle$ arising from the quantization of (single-mode) electromagnetic fields are not fully coherent for $n\neq 0$. Namely, with exception of $\vert n=0 \rangle$, the states $\vert n\geq 1 \rangle$ are not eigenvectors of the boson-annihilation operator. It may be proved that $\vert n=1 \rangle$ is first-order coherent, but it lacks second and higher-order coherence. The states $\vert n \geq 2 \rangle$ are also first-order coherent but they do not factorize the second-order correlation function. Then, the number states $\vert n \geq 1 \rangle$ are {\em nonclassical} in the sense that they are not fully coherent, so that no Maxwell field can be attached to them, and no classical description is possible. However, the `classical' property of states $\vert n \geq 1 \rangle$ to be first-order coherent justifies their  recurrent use as the incoming signal in contemporary versions of the Young's experiment \cite{Gra86,Rue13,Asp16}. On the other hand, the vacuum state $\vert n=0 \rangle$ belongs to the eigenvalue $0$ of the annihilation operator. As such eigenvalue is the trivial solution of the Maxwell equations, the zero-photon state $\vert n=0 \rangle$ corresponds to the classical notion of the absence of any field.

Despite the above remarks, the marvelous properties of quantum systems offer the possibility of using linear combinations of number states (rather than a given number state alone) to represent nontrivial  eigenvectors of the boson-annihilation operator. Denoting by $\vert \alpha \rangle$ one of such vectors, the square modulus $\vert \alpha \vert^2$ of the complex eigenvalue $\alpha$ provides the expectation value of the number of photons in the superposition. In turn, the real and imaginary parts of $\alpha$ supply the expectation values of the field-quadratures $\langle \hat x_1 \rangle = {\sqrt 2} \mbox{Re}(\alpha)$, $\langle \hat x_2 \rangle = {\sqrt 2} \mbox{Im}(\alpha)$. Therefore, the variances are equal $(\Delta \hat x_1)^2 = (\Delta \hat x_2)^2 = \sfrac12$ and the related uncertainty is minimized $\Delta \hat x_1 \Delta \hat x_2 = \sfrac12$. In other words, the vectors $\vert \alpha \rangle$ represent the closest quantum states to the Maxwell description of single-mode electromagnetic radiation (similar conclusions hold for the multi-mode case). A very important feature of the set $\{ \vert \alpha \rangle \}$ is that, although it is not orthogonal, this satisfies the resolution of the identity \cite{Kla60}. Thus, $\{ \vert \alpha \rangle \}$ is an  overcomplete basis of states for the quantized single-mode electromagnetic fields. This property was used by  Glauber \cite{Gla63c} and Sudarshan \cite{Sud63} to introduce a criterion to classify the fields as those that admit a classical description (like the fully coherent states) and the ones which are nonclassical (like the number states $\vert n \geq 1 \rangle$). The former can be written as a mixture of pure states $\vert \alpha \rangle \langle \alpha \vert$ for which the weights $P(\alpha)$ are admissible as conventional probabilities, the Dirac delta distribution $P(\alpha) = \delta(\alpha)$ included. The nonclassical fields are such that $P(\alpha)$ is not a conventional probability. 

Over the time, states other than $\vert \alpha \rangle$ were found to minimize the quadrature-uncertainty \cite{Tak65,Rob65,Sto70a,Sto71,Lu71,Lu72,Yue76,Dod80,Raj82,Yue83}. In contraposition with $\vert \alpha \rangle$, such states lead to $(\Delta \hat x_1)^2 \neq (\Delta \hat x_2)^2$ and, depending on a complex parameter $\xi$, one of the quadrature-variances can be squeezed by preserving the product $\Delta \hat x_1 \Delta \hat x_2 = \sfrac12$. Accordingly, the complementary variance is stretched. These properties found immediate applications in optical communication \cite{Yue78,Sha79,Yue80} and interferometry \cite{Har03,Sil08}, including the detection of gravitational waves \cite{Hol79,Bra80,Cav80a,Cav80b,Tho94,Sch10,Sau17}. The $\xi$--parameterized minimal uncertainty states $\vert \alpha, \xi \rangle$ are called {\em squeezed}  \cite{Hol79} (see also  \cite{Wal83,Lou87,Tei89}) and, like the number states $\vert n \geq 1 \rangle$, they admit no description in terms of the Maxwell theory. That is, the squeezed states $\vert \alpha, \xi \rangle$  are nonclassical.

As it can be seen, we have three different basis sets to represent the quantum states of single-mode (and multi-mode) electromagnetic radiation. Namely, the number states $\vert n \rangle$, the fully coherent states (hereafter {\em coherent states} for short) $\vert \alpha \rangle$, and the squeezed states $\vert \alpha, \xi \rangle$. The former and last states (with exception of $\vert n=0 \rangle$) are nonclassical while the coherent states may be described within the Maxwell theory. Depending on the optical field under study, we can use either of the above basis to make predictions and to explain experimental outcomes. Pretty interestingly, the classicalness of a given state is not invariant under linear superpositions. As immediate example recall that the `classical' state $\vert \alpha \rangle$ is a superposition of the nonclassical number states $\vert n \rangle$. In turn, it may be shown that nonclassical states $\vert \alpha, \xi \rangle$ can be expressed as a superposition of coherent states $\vert \alpha \rangle$ \cite{Dod74,Ger93,Mat96a,Roy98}. The `mystery' is hidden in the relative phases occurring as a consequence of any superposition of quantum states. According to Dirac, the reason because people had not thought of quantum mechanics much earlier is that the phase quantity was very well hidden in nature \cite{Moo89}, p.~218. Indeed, it is the probability amplitude of the entire superposition which expresses the difference between quantum and classical behavior. Thus, in quantum mechanics, not probabilities but probability amplitudes are summed up to make predictions! In the Young's experiment discussed above, for example, the sinusoidal modulation is the result of calculating the complete probability amplitude $\psi = \psi_1 + \psi_2$, with $\psi_1$ and $\psi_2$ the amplitudes relative to either pinhole. The modulation term $2 \mbox{Re}(\psi_1 \psi_2^*)$ of the entire probability $\vert \psi \vert^2$ is different from zero only when we have no information about the pinhole that actually emitted the detected photon. It is then relevant to find a form to measure classicalness in quantum states \cite{Ger93,Wig32,Cur14,Ken04,Aga13,Man79,Kim02,Wan02}. Besides the Glauber--Sudarshan $P$--representation \cite{Gla63c,Sud63}, the main criteria include the negativity of the Wigner function \cite{Wig32,Cur14,Ken04}, some asymmetries in the Wigner and other pseudo-probability distributions \cite{Aga13}, the identification of sub--Poissonian statistics (Mandel parameter) \cite{Man79}, and the presence of entanglement in the outcomes of a beam splitter (Knight conjecture) \cite{Kim02,Wan02}. 

The nonclassical properties of light have received a great deal of attention in recent years, mainly in connection with quantum optics \cite{Dod03}, quantum information \cite{Bou00}, and the principles of quantum mechanics \cite{Sil08}. Pure states representing fields occupied by a finite number of photons $n\neq 0$ exhibit nonclassical properties. The same holds for squeezed states and any other field state having sub-Poissonian statistics \cite{Man79}. Using some deformations of the algebra generated by the boson operators, other states have been constructed to represent photons with ``unusual properties'' \cite{Bie89,Mac89}, which may be applied in photon counting statistics, squeezing, and signal-to-quantum noise ratio \cite{Sol94}. Immediate generalizations \cite{Mat96b,Man97} motivated the development of the subject as an important branch of quantum optics \cite{Dod03}. Other deformations of the boson algebra include supersymmetric structures \cite{Coo95,Bag01,Aoy01a,Aoy01b,Mie04,And12,Gan18,Mie84,Bec88,Ele97,Ses98,Wil18,Ber93} for which the so-called polynomial Heisenberg algebras are quite natural \cite{Mie84,Fer94,Fer95,Ros96,Kum96,Fer99,Fer07,Fer08}. Recently, some non-Hermitian models have been shown to obey the distortions of the boson algebra that arise in the conventional supersymmetric approaches \cite{Ros15,Jai17,Bla18,Zel17b,Ros18b}. The deformed oscillator algebras have been used to construct the corresponding generalized (also called {\em nonlinear}) coherent states \cite{Mat96b,Man97,Ber93,Fer94,Fer95,Ros96,Kum96,Fer07,Fer99,Fer08,Zel17b,Ros18b,Orz88,Jun99,Roy00a,Rok04a,Rok04b,Tav08,Bag08,Twa08,Bag09a,Bag09b,Hon09,Abb09,Tav10,Saf11,Kor13,Noo14,Hus11,Ang12,Ang13,Zel17a,Moj18}. Most of these states exhibit nonclassical properties that distinguish them from the coherent states of the conventional boson algebra. 

The aim of the present work is to provide materials addressed to introduce the subject of coherent and squeezed states. The contents have been prepared for nonspecialists, so particular attention is given to the basic concepts as well as to their historical development. I preliminary apologize many authors because I have surely missed some fundamental references. In Section~\ref{basics} the fundamentals of optical detection and coherence are revisited. The meaning of the affirmation that the photon ``interferes only with itself'' and that two different photons cannot interfere is analyzed in detail (see Section~\ref{self}). The conditions for fully coherence are then given for fields of any number of modes and polarization. In Section~\ref{cs} the coherent states for single-mode field are analyzed at the time that their fundamental properties are revisited. Spacial attention is addressed to the wave packets of minimum uncertainty that can be constructed for the conventional oscillator  (Sections~\ref{owp} and \ref{swp}), where the historical development of related ideas and concepts is overview. Some generalizations for wave packets with widths that depend on time are given in Section~\ref{twp}. The discussion about the wave packets for the hydrogen atom, as well as its historical development and controversies, is given in Sections~\ref{portrait} and \ref{controversy}. In Section~\ref{groupw} the connection with representations of Lie groups in terms of generalized coherent states is reviewed in introductory form. The notions of generalized coherent states are discussed in Section~\ref{gral}, some analogies with classical systems are also indicated. Final comments are given in Section~\ref{concluye}.

%----------------------------------------------------------->
\section{Basics of quantum optical detection and coherence}
\label{basics}

The quantized electric field is represented by the Hermitian operator 
\be
{\mathbf E} ({\mathbf r}, t) = {\mathbf E}^{(+)} ({\mathbf r}, t) + {\mathbf E}^{(-)} ({\mathbf r}, t), 
\label{field}
\ee
where its positive and negative frequency parts, ${\mathbf E}^{(+)} ({\mathbf r}, t)$ and ${\mathbf E}^{(-)} ({\mathbf r}, t)$, are mutually adjoint
\be
{\mathbf E}^{(+)} ({\mathbf r}, t)  = {\mathbf E}^{(-) \dagger} ({\mathbf r}, t).
\label{masm}
\ee
Details concerning field quantization can be consulted in e.g. \cite{Man95}. The positive frequency part ${\mathbf E}^{(+)} ({\mathbf r}, t)$ is a photon annihilation operator \cite{Gla63b}, so it is bounded from below ${\mathbf E}^{(+)} ({\mathbf r}, t) \vert \mbox{vac} \rangle =0$, with $\vert \mbox{vac} \rangle$ the state in which the field is empty of all photons. In turn, ${\mathbf E}^{(-)} ({\mathbf r}, t)$ is a photon creation operator, with no upper bound. In particular, ${\mathbf E}^{(-)} ({\mathbf r}, t) \vert \mbox{vac} \rangle$ represents a one-photon state of the field. 

Following Glauber \cite{Gla07}, let us associate the action of an ideal photodetector with the operator ${\mathbf E}^{(+)} ({\mathbf r}, t)$. Assuming that the field is in state $\vert i \rangle$, and that one photon (polarized in the $\mu$-direction) has been absorbed, after the photo-detection we have ${\mathbf E}^{(+)}_{\mu} ({\mathbf r}, t) \vert i \rangle$. The probability that such a result coincides with the state $\vert f \rangle$ is regulated by the probability amplitude 
\be
{\cal A}_{i \rightarrow f}^{(1)} = \langle f \vert {\mathbf E}^{(+)}_{\mu} ({\mathbf r}, t) \vert i \rangle,
\ee
which is a complex number in general. Notice that we do not require to know which of the possible states of the field is $\vert f \rangle$. The only requirement is that $\vert f \rangle$ be a physically admissible state. Then
\be
{\cal P}_{i \rightarrow f}^{(1)} = \vert {\cal A}_{i \rightarrow f}^{(1)}  \vert^2 = \langle i \vert {\mathbf E}^{(-)}_{\mu} ({\mathbf r}, t) \vert f \rangle \langle f \vert {\mathbf E}^{(+)}_{\mu} ({\mathbf r}, t) \vert i \rangle
\ee
is the probability we are looking for. To obtain the probability per unit time ${\cal P}_{det}^{(1)}$ that an individual photon be absorbed by the ideal detector at point $\mathbf r$ at time $t$, we have to sum over all possible (admissible) states
\be
{\cal P}_{det}^{(1)} (\mathbf r, t) = \sum_f {\cal P}_{i \rightarrow f}^{(1)} = \langle i \vert {\mathbf E}^{(-)}_{\mu} ({\mathbf r}, t) \left[  \sum_f \vert f \rangle \langle f \vert 
\right]{\mathbf E}^{(+)}_{\mu} ({\mathbf r}, t) \vert i \rangle.
\label{casi}
\ee
Now, it is quite natural to assume that the (admissible) final states form a complete orthonormal set. Therefore, the sum of projector operators $\vert f \rangle \langle f \vert$ between brackets in (\ref{casi}) can be substituted by the identity operator $\mathbb I$ to get
\be
{\cal P}_{det}^{(1)} (\mathbf r, t) = \langle i \vert {\mathbf E}^{(-)}_{\mu} ({\mathbf r}, t) {\mathbf E}^{(+)}_{\mu} ({\mathbf r}, t) \vert i \rangle = \vert\vert {\mathbf E}^{(+)}_{\mu} ({\mathbf r}, t) \vert i \rangle \vert\vert^2.
\label{casi2}
\ee
That is, probability ${\cal P}_{det}^{(1)}$ coincides with the expectation value of the Hermitian product ${\mathbf E}^{(-)}_{\mu} ({\mathbf r}, t) {\mathbf E}^{(+)}_{\mu} ({\mathbf r}, t) $, evaluated at the initial state $\vert i \rangle$ of the field. Equivalently, this is equal to the square norm of the vector ${\mathbf E}^{(+)}_{\mu} ({\mathbf r}, t) \vert i \rangle$, which represents the state of the field just after the action of the ideal detector. These results show that the detector we have in mind measures the average value of the product ${\mathbf E}^{(-)}_{\mu} ({\mathbf r}, t) {\mathbf E}^{(+)}_{\mu} ({\mathbf r}, t) $, and not the average of the square of the Hermitian operator (\ref{field}) representing the field \cite{Gla63b}. Thus, the field intensity $I$, as a quantum observable, is represented by the operator $\mathbf E^{(-)} \mathbf E^{(+)}$, and not by the operator $\mathbf E^2$. It is illustrative to rewrite (\ref{casi2}) as follows
\be
{\cal P}_{det}^{(1)} (\mathbf r, t) = \langle i \vert \hat I_{\mu} ({\mathbf r}, t)  \vert i \rangle, \quad \hat I ({\mathbf r}, t) ={\mathbf E}^{(-)}  ({\mathbf r}, t) {\mathbf E}^{(+)} ({\mathbf r}, t),
\label{casi3}
\ee
which makes evident that ${\cal P}_{det}^{(1)}$ is the expectation value of the intensity $\hat I_{\mu} ({\mathbf r}, t)$. Notice that $\vert i \rangle = \vert \mbox{vac} \rangle$ produces ${\cal P}_{det}^{(1)}=0$, as this would be expected. The above results can be easily extended to arbitrary initial states (either pure or mixed) represented by $\rho$ as follows
\be
{\cal P}_{det}^{(1)} (y) = \mbox{Tr} \left\{ \rho {\mathbf E}^{(-)}_{\mu} (y) {\mathbf E}^{(+)}_{\mu} (y)  \right\}, \quad y \equiv  (\mathbf r, t).
\ee
A lucky researcher has at his disposal more than one photodetector in his laboratory. He can use two detectors situated at different space--time points $y_1$ and $y_2$ to detect photon (delayed) coincidences. The probability amplitude associated with his predictions is of the form
\be
{\cal A}_{i \rightarrow f}^{(2)} = \langle f \vert {\mathbf E}^{(+)}_{\mu} (y_2) {\mathbf E}^{(+)}_{\mu} (y_1) \vert i \rangle,
\ee
so the probability per unit $(\mbox{time})^2$ that one photon is recorded at $y_1$ and another at $y_2$ is given by the expression
\be
{\cal P}_{det}^{(2)} (y_1,y_2) = \mbox{Tr} \left\{ \rho {\mathbf E}^{(-)}_{\mu} (y_1) {\mathbf E}^{(-)}_{\mu} (y_2) {\mathbf E}^{(+)}_{\mu} (y_2) {\mathbf E}^{(+)}_{\mu} (y_1)  \right\}.
\label{p2}
\ee
To rewrite (\ref{p2}) in terms of the intensity operator $\hat I (y)$, it is customary to use the normally ordered notation
\be
:{\mathbf E}^{(-)} {\mathbf E}^{(+)}  {\mathbf E}^{(-)} {\mathbf E}^{(+)} :
\,\, = \,\, {\mathbf E}^{(-)} {\mathbf E}^{(-)} {\mathbf E}^{(+)} {\mathbf E}^{(+)}.
\label{normal}
\ee 
Therefore
\be
{\cal P}_{det}^{(2)} (y_1,y_2) = \langle i \vert \, :\hat I_{\mu} (y_2) \hat I_{\mu} (y_1) : \, \vert i \rangle
\ee
corresponds to the expectation value of the square-intensity observable, which formalizes the experimental outcomes obtained by Brown and Twiss \cite{Wal79}. 

The above results can be generalized at will to include an arbitrary number of photodetectors (I have in mind a researcher even more fortunate than the previous one!). Another generalization may be addressed to investigate the correlations of the fields at separated positions and times. In this context, Glauber introduced the first-order correlation function
\be
G^{(1)} (y_1,y_2) = \mbox{Tr} \left\{ \rho {\mathbf E}^{(-)}_{\mu} (y_1) {\mathbf E}^{(+)}_{\mu} (y_2)  \right\},
\ee
which is complex-valued in general and satisfies $G^{(1)} (y,y) = {\cal P}_{det}^{(1)} (y)$. The expression for the $n$-th-order correlation function
\be
G^{(n)} (y_1, \ldots, y_n, y_{n+1}, \ldots, y_{2n}) = \mbox{Tr} \left\{
\rho \prod_{k=1}^n {\mathbf E}^{(-)}_{\mu} (y_k) 
\prod_{\ell =n+1}^{2n} {\mathbf E}^{(+)}_{\mu} (y_{\ell})
\right\}
\label{G}
\ee
is now clear. The normalized form of the above formula is defined as
\be
g^{(n)} (y_1, \ldots, y_{2n}) = \frac{ G^{(n)} (y_1, \ldots, y_{2n}) }{
\prod_{k=1}^{2n} \left\{ G^{(1)} (y_k, y_k) \right\}^{1/2} }
\equiv \frac{ G^{(n)} (y_1, \ldots, y_{2n}) }{
\prod_{k=1}^{2n} \left\{ {\cal P}_{det}^{(1)} (y_k)  \right\}^{1/2} }.
\label{Gn}
\ee
Thus, $g^{(n)}$ is the $n$-th-order correlation function $G^{(n)}$, weighted by the root-squared product of the probabilities that one photon is detected at $y_1$, another at $y_2$, and so on until all the $2n$ space-time points $y_k$ have been exhausted. Notice that the product of probabilities ${\cal P}_{det}^{(1)} (y_k)$ means independence in detecting the individual photons. 

Glauber found that $\vert g^{(n)} \vert =1$, $n=1,2, \ldots$, is a necessary condition for coherence. The simplest way to satisfy such a requirement is by demanding the factorization of $G^{(n)}$ as follows
\be
\left\vert G^{(n)} (y_1, \ldots, y_{2n}) \right\vert = \prod_{k=1}^{2n} \left\{ G^{(1)} (y_k, y_k) \right\}^{1/2}.
\label{factor}
\ee
In other words, if the correlations of a given field at $2n$ space-time points $y_k$ can be expressed, up to a phase, as the root-squared product of the one-photon detection probabilities ${\cal P}_{det}^{(1)} (y_k)$, then the field is $n$-th-order coherent. If the latter condition is fulfilled for all orders the field is fully coherent. General properties of the functions $G^{(n)}$ and $g^{(n)}$ can be consulted in \cite{Gla07}.

%----------------------------------------------------------->
\subsection{Self-interference of single photons}
\label{self}

To provide an immediate example let us compare the Glauber's first-order correlation function $g^{(1)}(y_1,y_2)$ with its counterpart $\lambda_{(1,2)}$ in the Young's experiment. As a first conclusion we have $G^{(1)}(y_k,y_k) = {\cal P}^{(1)}_{det}(y_k) = \langle \hat I (y_k) \rangle$, $k=1,2$, which verifies that the Hermitian operator ${\mathbf E}^{(-)} {\mathbf E}^{(+)}$ represents the `quantum observable' of field intensity $\hat I$. Second, the normalization condition $\vert g^{(1)}(y_1,y_2) \vert =1$ means $g^{(1)}(y_1,y_2) = \exp[i \varphi (y_1,y_2) ]$, so that $\varphi (y_1,y_2)  = \theta_{(1,2)}$. Using these results the visibility (\ref{young}) can be rewritten in the form
\be
{\cal V}_Y = \frac{ 2 \sqrt{ G^{(1)}(y_1,y_1) G^{(1)}(y_2,y_2)}
}{G^{(1)}(y_1,y_1) + G^{(1)}(y_2,y_2) } \left\vert g^{(1)}(y_1,y_2) \right\vert,
\label{young2}
\ee
which is simplified to 
\be
{\cal V}_Y = \frac{ 2 \sqrt{ G^{(1)}(y_1,y_1) G^{(1)}(y_2,y_2)} }{G^{(1)}(y_1,y_1) + G^{(1)}(y_2,y_2) } 
\label{young3}
\ee
for first-order coherent fields (i.e., if $\vert g^{(1)}(y_1,y_2) \vert =1$). In such case, the Young's visibility (\ref{young3}) is equal to 1 whenever $G^{(1)}(y_1,y_1) = G^{(1)}(y_2,y_2)$, which is equivalent to ${\cal P}^{(1)}_{det}(y_1) = {\cal P}^{(1)}_{det}(y_2)$. On the other hand, given ${\cal P}^{(1)}_{det}(y_k) = \langle \hat I (y_k) \rangle$, we may interpret ${\cal P}^{(1)}_{det}(y_k)$ as the probability that one photon emitted from the $k$-th pinhole has been recorded by the detector. In this sense the identity ${\cal P}^{(1)}_{det}(y_1) = {\cal P}^{(1)}_{det}(y_2)$ means that we cannot determine which of the two pinholes is the one that emitted such a photon. Thus, in the Young's experiment for a first-order coherent field, if the detection of an individual photon implies lack of knowledge about the source, interference fringes will be produced with maximum visibility. Our affirmation is particularly relevant for a single-photon wave-packet that impinges on the Young's interferometer. As ``any pure state in which the field is occupied by a single photon possesses first order coherence'' \cite{Gla07}, p.~62, the single-photon wave-packet is able to produce interference fringes with maximum visibility, so this may be classified as highly coherent in the ordinary sense. 

We would like to emphasize that, although the above results might be put in correspondence with the very famous sentence of Dirac that ``each photon then interferes only with itself. Interference between two different photons can never occur'', we must take it with a grain of salt. On the one hand, the origin of the sentence can be traced back to the first edition of the Dirac's book, published in 1930 \cite{Dir30}, long before sources of coherent light like the maser (1953) or the laser (1960) were built. Then,  by necessity, Dirac used the ordinary notion of coherence to formulate his sentence. The correlations of photons discussed in \cite{Dir30}, Ch.~I, are thus of the first-order in the Glauber sense. In other words, the phrase  ``each photon then interferes only with itself'' applies to conventional interferometry only (in the Young's experiment, for example). On the other hand, although the Brown--Twiss results and the Glauber theory were published much later than the first edition of the Dirac's book, it is also true that the sentence we are dealing with survived, with minimal modifications, until the fourth edition (revised) of the book, published in 1967 \cite{Dir67}. Therefore, it seems that even in 1967 Dirac was not aware that single-photon fields lack second- and higher-order coherence. Other option is that he was not interested in making the appropriate adjustments to his manuscript. In my opinion, the latter option is in opposition to the Dirac's perfectionism, so it can be discarded. The former option is viable but unlikely for  somebody as learned as Dirac. A third option is that Dirac was aware of the Brown--Twiss and Glauber works but considered them as not definitive. To me, this last is the most reasonable since many people was reluctant to accept the Brown--Twiss results \cite{Bro74}. Besides, the Glauber theory, albeit corroborated on the blackboard, was far from being experimentally confirmed at the time. In any case, phenomena associated to the second-order correlation function (including the Brown--Twiss effect) are experimentally observed over and over in quantum-optics laboratories around the world. So the second part of Dirac's sentence ``interference between two different photons can never occur'' is also currently defeated. Other remarks in the same direction can be found in \cite{Jay80}.

Nevertheless, it is remarkable that efforts to produce interference with `feeble light' were reported as early as 1905 by Taylor \cite{Tay09}. Fundamental advances on the single-photon interference arrived up to 1986, with the experimental results of Grangier, Roger and Aspect about the anticorrelation effect produced on individual photons by a beam splitter \cite{Gra86}.

%----------------------------------------------------------->
\subsection{Fully coherent states of quantized radiation fields}

The touchstone used by Glauber to determine the quantum states that satisfy the factorization property (\ref{factor}) is reduced to the eigenvalue equation 
\be
{\mathbf E}^{(+)}_{\mu} (y) \vert \mbox{\bf ?} \rangle = {\cal E}_{\mu} (y) \vert \mbox{\bf ?} \rangle.
\label{eigen}
\ee
That is, the states $\vert \mbox{\bf ?} \rangle$ which Glauber was looking for should be eigenvectors of the positive frequency operator ${\mathbf E}^{(+)}_{\mu} (y)$. As the latter is not self-adjoint, two features of the solutions to (\ref{eigen}) are easily recognized. First, the eigenvalues ${\cal E}_{\mu} (y)$ are complex numbers in general. Second, the (possible) orthogonality of the set of  solutions $\vert \mbox{\bf ?} \rangle$ is not (automatically) granted. Nevertheless, assuming (\ref{eigen}) is fulfilled, the introduction of the state $\rho = \vert \mbox{\bf ?} \rangle \langle \mbox{\bf ?} \vert$ into (\ref{G}) gives
\be
G^{(n)} (y_1, \ldots, y_{2n}) = 
\prod_{k=1}^n {\cal E}^*_{\mu} (y_k) 
\prod_{\ell =n+1}^{2n} {\cal E}_{\mu} (y_{\ell}),
\ee
where $z^*$ stands for the complex-conjugate of $z \in \mathbb C$. Clearly, the above expression satisfies (\ref{factor}) and produces $\vert g^{(n)} \vert =1$. Thus, the eigenvectors of ${\mathbf E}^{(+)}_{\mu} (y)$ belonging to complex eigenvalues are the most suitable to represent fully coherent states.

%----------------------------------------------------------->
\section{Single-mode coherent states}
\label{cs}

The electric-field operator (\ref{field}) for a single-mode of frequency $\omega$, linearly polarized in the $x$-direction, with $z$-spatial dependence, can be written as
\be
E(z,t) =  {\cal E}_F \left[ \frac{a(t) + a^{\dagger}(t)}{\sqrt{2}} \right], \quad {\cal E}_F =  {\cal E}_{vac} \sin (kz),
\label{single}
\ee
where $k=\omega/c$ is the wave-vector and ${\cal E}_{vac}$ (expressed in electric field units) is a measure of the minimum size of the quantum optical noise that is inherent to the field \cite{Lou87}. The latter is associated to the vacuum fluctuations of the field since it is the same for any strength of excitation (even in the absence of any excitation) of the mode. The mutually adjoint time--dependent operators, $a(t)$ and $a^{\dagger}(t)$, are defined in terms of the boson ladder operators $[a,a^{\dagger}]=1$ as usual $a(t) = a \exp(-i\omega t)$. It is useful to introduce the field quadratures
\be
\hat x_1 =\frac{1}{\sqrt 2} (a^{\dagger} + a), \quad \hat x_2 =\frac{i}{\sqrt 2} (a^{\dagger} - a),  \quad [\hat x_1, \hat x_2]=i,
\ee
to write
\be
E(z,t) = {\cal E}_F (\hat x_1 \cos \omega t + \hat x_2 \sin \omega t).
\ee
At $t=0$ we have $E(z,0) = {\cal E}_F \hat x_1$. That is, the quadrature $\hat x_1$ represents the (initial) electric field. It is not difficult to show that the conjugate quadrature $\hat x_2$ corresponds to the (initial) magnetic field \cite{Lou87}.

Equipped with physical units of position and momentum, the quadratures $\hat x_1$ and $\hat x_2$ can be put in correspondence with a pair of phase--space operators 
\be
\hat x_1 = \sqrt{\frac{m \omega}{\hbar} } \hat q, \quad \hat x_2 = \frac{1}{\sqrt{m \hbar \omega}} \hat p, \quad [\hat q, \hat p] = i\hbar,
\label{osc1}
\ee
so they define the oscillator-like Hamiltonian $\hat H =\hbar \omega H$, with $H$ the dimensionless quadratic operator
\be
H = \frac12 \left( \hat x_1^2 + \hat x_2^2 \right) = \frac12 \left( \frac{m\omega}{\hbar} \hat q^2 + \frac{1}{m\hbar \omega} \hat p^2 \right).
\label{osc}
\ee

%----------------------------------------------------------->
\subsection{Field correlations}

Assuming we are interested in photon delayed coincidences, we may compute field correlations at $t$ and $t + \tau$, measured at the same space point. Introducing (\ref{single}) into  ({\ref{Gn}) we obtain
\be
g^{(1)}(t,t+\tau)= e^{i\omega \tau},
\label{corr1}
\ee
and
\be
g^{(2)} (t,t+\tau)= \frac{ \langle i \vert a^{\dagger} a^{\dagger} aa \vert i \rangle }{ \langle i \vert a^{\dagger} a \vert i \rangle^2 }  = 1 - \frac{1}{\langle i \vert \hat n \vert i \rangle}, \quad \langle i \vert \hat n \vert i \rangle \neq 0,
\label{corr2}
\ee
where we have used the photon-number operator $a^{\dagger} a =\hat n$.  Notice that $g^{(1)}$ depends on the delay $\tau$ between detections and not on the initial time $t$, so we write $g^{(1)}(t,t+\tau) =g^{(1)}(\tau)$. Besides, $g^{(2)}$ does not depend on any time variable, so we can write $g^{(2)} (t,t+\tau) = g^{(2)} (0)$. 

The first-order correlation function (\ref{corr1}) shows that any field represented by the operator (\ref{single}) is first-order coherent. That is, any quantized single-mode field $E(z,t)$ is coherent in the ordinary sense! In particular, the single-photon fields thought by Dirac in his book belong to the class $g^{(1)} =  \exp(i \omega \tau)$, see our discussion on the matter in Section~\ref{self}. In turn, the second-order correlation function (\ref{corr2}) shows that the single-photon fields $\vert i \rangle = \vert n=1 \rangle$ produce the trivial result $g^{(2)} =0$, so they lack second- and higher order coherence, as we have already indicated. On the other hand, the field occupied by two or more photons $\vert i \rangle = \vert n \geq 2 \rangle$ lead to $g^{(2)} = 1 -1/n$. As the state $\vert n \geq 2 \rangle$ produce $g^{(2)} < 1$ for a finite number of photons, we know that the factorization (\ref{factor}) is not admissible if $n \geq 2$. Of course, $g^{(2)} \rightarrow 1$ as $n \rightarrow \infty$. Note also that the probabilities of detecting a photon at time $t$, and another one at time $t + \tau$, produce the same result ${\cal P}^{(1)}_{det}(t) = {\cal P}^{(1)}_{det}(t + \tau) = \vert {\cal E}_F \vert^2 n/2$, $n=2,3,\ldots$, which does not depend on any time variable.

%----------------------------------------------------------->
\subsection{Mandel parameter}

To get additional information about the initial state $\vert i \rangle$ of the field, let us rewrite (\ref{corr2}) as follows
\be
g^{(2)} (0) = 1 + \frac{ (\Delta \hat n)^2 - \langle i \vert \hat n \vert i \rangle}{ \langle i \vert \hat n \vert i \rangle^2},
\label{corr3}
\ee
where we have added a zero to complete the photon-number variance $(\Delta \hat n)^2 = \langle i \vert \hat n^2 \vert i \rangle - \langle i \vert \hat n \vert i \rangle^2$. Introducing the Mandel parameter \cite{Man79}:
\be
Q_M= \frac{ (\Delta \hat n)^2 - \langle i \vert \hat n \vert i \rangle}{ \langle i \vert \hat n \vert i \rangle}
 =  \frac{ (\Delta \hat n)^2 }{ \langle i \vert \hat n \vert i \rangle} -1, \quad \langle i \vert \hat n \vert i \rangle \neq 0,
\label{mandel}
\ee
we arrive at the relationships
\be
g^{(2)} (0)= 1 + \frac{Q_M}{ \langle i \vert \hat n \vert i \rangle}, \quad Q_M= \left[ g^{(2)} (0) - 1 \right]  \langle i \vert \hat n \vert i \rangle, \quad \quad \langle i \vert \hat n \vert i \rangle \neq 0.
\label{corr4}
\ee
For any number state $\vert n \rangle$ we clearly have $(\Delta \hat n)^2 =0$, so that $g^{(2)} = 1 -1/n$ and $Q_M=-1$ for $n \geq 1$. 

%----------------------------------------------------------->
\subsection{Klauder-Glauber-Sudarshan states}
\label{KGS}

Using ${\cal E}_F$ as the natural unit of the electric field strength \cite{Lou87}, and dropping the phase-time dependence of $a(t)$, the introduction of  (\ref{single}) into the eigenvalue equation (\ref{eigen}) gives $a \vert \alpha \rangle = \alpha \vert \alpha \rangle$, the solution of which can be written as the normalized superposition 
\be
\vert \alpha \rangle = e^{-\vert \alpha \vert^2/2} \sum_{n=0}^{\infty} \frac{\alpha^n}{\sqrt{n!}} \vert n \rangle, \quad \alpha \in \mathbb C.
\label{glauber1}
\ee
The above result was reported by Glauber in his quantum theory of optical coherence \cite{Gla07}. Remarkably, the superposition (\ref{glauber1}) was previously used (in implicit form) by Schwinger in his studies on quantum electrodynamics \cite{Sch53}, and introduced by Klauder as the generator of an {\em overcomplete family of states} which is very appropriate to study the Feynman quantization of the harmonic oscillator \cite{Kla60}. Indeed, Klauder realized that $\langle \alpha \vert a^{\dagger} a \vert \alpha \rangle = \alpha^* \alpha$, and proved that the set of states $\vert \alpha \rangle$ forms a ``basis'' for the oscillator's Hilbert space 
\be
\frac{1}{\pi} \int d^2 \alpha \vert \alpha \rangle \langle \alpha \vert = \sum_{n=0}^{\infty} \vert n \rangle \langle n \vert = \mathbb I.
\label{identity}
\ee
Notwithstanding, the basis elements are not mutually orthogonal
\be
\langle \beta \vert \alpha \rangle= \exp \left[ - \frac{\vert \alpha - \beta \vert^2}{2} + i \mbox{Im}( \beta^* \alpha)\right], \quad \vert \langle \beta \vert \alpha \rangle \vert^2 = e^{-\vert \alpha - \beta \vert^2}.
\ee
Further improvements of the mathematical structure associated to the superpositions (\ref{glauber1}) were provided by Klauder himself in his continuous representation theory \cite{Kla63a,Kla63b}. In addition to the Glauber contributions \cite{Gla07}, fundamental properties of these states addressed to the characterization of light beams were also discussed by Sudarshan \cite{Sud63}. Hereafter the superpositions (\ref{glauber1}) will be called `Klauder-Glauber-Sudarshan states' or `{\em canonical} coherent states' (KGS-states or coherent states for short).

The probability that the field represented by $\vert \alpha \rangle$ is occupied by $n$-photons is given by the Poisson distribution:
\be
{\cal P}_{\alpha \rightarrow n} = \vert \langle n \vert \alpha \rangle \vert^2 =\frac{ (\overline n)^n e^{-\overline n }}{n!},
\label{poisson}
\ee
where $\overline n \equiv \langle \alpha \vert \hat n \vert \alpha \rangle =\vert \alpha \vert^2$. As it is well known, Poisson distributions like (\ref{poisson}) are useful for describing random events that occur at some known average rate. In the present case, the rate is ruled by the mean value of the photon-number $\overline n$. Mandel and Wolf provided a very pretty physical interpretation: ``when light from a single-mode laser falls on a photodetector, photoelectric pulses are produced at random at an average rate proportional to the mean light intensity, and the number of pulses emitted within a given time interval therefore obeys a Poisson distribution'' \cite{Man95}, pp. 23-24. In addition, the distribution we are dealing with is characterized by the fact that the variance $(\Delta \hat n)^2$ is equal to the mean value $\langle \alpha \vert \hat n \vert \alpha \rangle$, which can be easily verified. Therefore, if the initial state of the field in the relationships (\ref{corr4}) is a KGS-state $\vert \alpha \rangle$, then $(\Delta \hat n)^2= \langle \alpha \vert \hat n \vert \alpha \rangle =\vert \alpha \vert^2$, and $Q_M=0$. This result is quite natural by noticing that the normalization condition $\vert g^{(2)} \vert =1$ is automatically fulfilled by $\vert \alpha \rangle$.

On the other hand, the straightforward calculation shows that: 

(I) States $\vert \alpha \rangle$ evolve in time by preserving their form (i.e., they have temporal stability):
\be
\vert \alpha (t) \rangle = e^{-iHt} \vert \alpha \rangle = e^{-i\omega t/2} \vert \alpha e^{-i\omega t} \rangle.
\label{alphat}
\ee

(II) They are displaced versions of the vacuum state, $\vert \alpha \rangle=D(\alpha) \vert n=0 \rangle$, with
\be
D(\alpha) = \exp \left( \alpha a^{\dagger} - \alpha^* a \right) = \exp \left( -\vert \alpha \vert^2 \right) \exp \left( \alpha a^{\dagger} \right) \exp \left( -\alpha^* a \right)
\label{d1}
\ee
the unitary displacement operator fulfilling
\be
D(\alpha) a D^{\dagger} (\alpha) = a -\alpha, \quad D^{\dagger} (\alpha) a^{\dagger}  D(\alpha) = a^{\dagger}  -\alpha^*.
\label{d2}
\ee

(III) They are such that $\langle \hat x_1 \rangle = {\sqrt 2} \mbox{Re} (\alpha)$, $\langle \hat x_2 \rangle = {\sqrt 2} \mbox{Im} (\alpha)$, and $\langle \hat x_k^2 \rangle = \langle \hat x_k \rangle^2 +\frac12$, $k=1,2$. That is, the states $\vert \alpha \rangle$ minimize the uncertainty of quadratures
\be
( \Delta \hat x_1)^2 = ( \Delta \hat x_2)^2 =\frac12, \quad \Delta \hat x_1 \Delta \hat x_2= \frac12.
\label{osc2}
\ee

Any of the properties (I)--(III), including the eigenvalue equation $a \vert \alpha \rangle = \alpha \vert \alpha \rangle$ and the identity resolution (\ref{identity}), referred to as properties (A) and (B) in the sequel, can be assumed as the definition of the canonical coherent states. Then,  the remaining properties may be derived as a consequence. For systems other than the harmonic oscillator it is very well known that not all the properties (A), (B) and (I)--(III), are simultaneously satisfied. It is then customary to construct states by using either of the above properties and to call them {\em generalized coherent states} (see Section~\ref{gral}). If by chance the states so constructed exhibit any other property of the KGS-states, the `coherence criterion' can be refined in each case. Not all generalized coherent states are {\em classical} in the sense established by the Glauber theory, so they usually deserve a study of their classicalness to be classified. After all: coherent states are superpositions of basis elements to which some specific properties are requested on demand \cite{Gla94}.

%----------------------------------------------------------->
\subsection{Glauber-Sudarshan $P$- and Fock-Bargmann representations}

One of the most remarkable benefits offered by the KGS-states is the possibility of expressing any state of the radiation field as follows \cite{Gla63c,Sud63}:
\be
\rho = \int P(\alpha) \vert \alpha \rangle \langle \alpha \vert d^2 \alpha.
\label{P}
\ee
The above `diagonal' form of the density operator $\rho$ expresses the idea of having a mixed state, even though the basis defined by $\vert \alpha \rangle$ is not orthogonal. As the superposition of pure states $\vert \alpha \rangle \langle \alpha \vert$ defined above must be convex, the following condition is impossed
\be
\int P(\alpha) d^2 \alpha = 1.
\ee
Glauber introduced (\ref{P}) to study thermal fields \cite{Gla63c} and coined the term $P$-representation for it. In turn, Sudarshan argued that such representation is valid provided that $P(\alpha)$ is a conventional probability distribution \cite{Sud63}. It is easy to see that the state $\rho = \vert \beta \rangle \langle \beta \vert$, with $\vert \beta \rangle$ a GKS-state, implies $P(\alpha) = \delta (\alpha - \beta)$ by necessity. Then, the $P$-function are permitted to be as singular as the Dirac's delta distribution. It may be shown that the number states $\vert n \geq 1 \rangle$ are $P$-represented by the $n$-th derivatives of the delta distribution, so the latter is stronger singular than $\delta (z)$ and are not allowed as probabilities. Therefore, the states $\vert n \geq 1 \rangle$ admit no classical description in terms of the convex superposition (\ref{P}). For states other than the oscillator ones, the criterion applies in similar form. Details and general properties of the $P$-function can be consulted in the book by Klauder and Sudarshan \cite{Kla68}.

Using the identity (\ref{identity}), one can write any element $\vert \psi \rangle$ of the Hilbert space ${\cal H}$ as the superposition
\be
\vert \psi \rangle = \frac{1}{\pi} \int d^2 \alpha \psi(\alpha) \vert \alpha \rangle.
\ee
The Fourier coefficients $\psi(\alpha)$ are analytic over the whole complex $\alpha$-plane. Indeed, as these functions are holomorphic and are in one-to-one correspondence with the number eigenstates, they are elements of a Hilbert space of entire functions ${\cal F}$ named after Fock \cite{Foc28} and Bargmann \cite{Bar61}. The representation of the ladder operators $a$ and $a^{\dagger}$ in the ${\cal F}$--space corresponds to the derivative with respect to $\alpha$ and the multiplication by $\alpha$, respectively. The properties of the Fock-Bargmann functions $\psi(\alpha)$ and the related representations are studied by Saxon, under the name `creation operator representation', in his Book on quantum mechanics (1968) \cite{Sax68}, Ch.~VI.e (where no reference is given to neither Fock nor Bargmann works!).

%----------------------------------------------------------->
\subsection{Oscillator wave packets}
\label{owp}

Let us calculate the superposition (\ref{glauber1}) in the $x_1$-quadrature representation 
\be
\psi_{\alpha}(x) := \langle x \vert \alpha \rangle = e^{-\vert \alpha \vert^2/2} \sum_{n=0}^{\infty} \frac{\alpha^n}{\sqrt n!} \varphi_n(x),
\label{wave1}
\ee
where $\hat x_1 \vert x \rangle = x \vert x \rangle$, $x \in \mathbb R$, $\int_{\mathbb R} dx \vert x \rangle \langle x \vert = \mathbb I$, and
\be
\varphi_n(x) := \langle x \vert n\rangle = \frac{e^{-x^2/2}}{\pi^{1/4} (2^n n!)^{1/2}} H_n(x), \quad n=0,1,\ldots
\label{wave2}
\ee
The expression $H_n(x)$ stands for the Hermite-polynomials \cite{Abr64}, and $\varphi_n(x)$ represents the wave-functions of the harmonic oscillator \cite{Sax68,Sch49}. After introducing (\ref{wave2}) into (\ref{wave1}), and using the generating function of $H_n(x)$, Eq.~22.9.17 of Ref~\cite{Abr64}, we arrive at the Gaussian-like expression
\be
\psi_{\alpha}(x) = \pi^{-1/4} \exp \left[ - \frac{ (x - \langle \hat x_1 \rangle)^2}{2} + i \langle \hat x_2 \rangle (x  - \langle \hat x_1 \rangle) \right].
\label{wave3}
\ee
Using (\ref{osc2}) the above result acquires the familiar form of a wave packet
\be
\psi_{\alpha}(x) = \frac{1}{[2 \pi (\Delta \hat x_1)^2]^{1/4}} \exp \left[ - \frac{ (x - \langle \hat x_1 \rangle)^2}{4 (\Delta \hat x_1)^2} + i \langle \hat x_2 \rangle (x - \langle \hat x_1 \rangle) \right].
\label{wave32}
\ee
The Gaussian wave packet $\psi_{\alpha}(x)$ is localized about the point $x = \langle \hat x_1 \rangle = \sqrt 2 \mbox{Re} (\alpha)$, within a neighborhood defined by $\Delta \hat x_1 = 1/\sqrt 2$. Finally, equipped with physical units of position and momentum, the function (\ref{wave32}) is given by
\be
\psi_{\alpha}(x) = \left( \frac{\hbar}{m\omega} \right)^{1/4}
\frac{1}{[2 \pi (\Delta \hat q)^2]^{1/4} } \exp \left[ - \frac{ (q - \langle \hat q \rangle)^2}{4 (\Delta \hat q)^2} + i \frac{\langle \hat p \rangle (q - \langle \hat q \rangle)}{\hbar}
\right].
\label{packet}
\ee
For arbitrary values of $\Delta \hat q$, expression (\ref{packet}) coincides with the normalized {\em minimum wave packets} reported by Schiff in 1949 \cite{Sch49}. The smaller $\Delta \hat q$ is, the more localized the wave packet. Although Schiff derived (\ref{packet}) by thinking on the free particle motion, he was certain that ``the structure of this minimum packet is the same whether or not the particle is free, since this form can be regarded simply as the initial condition on the solution of the Schr\"odinger equation for any $V$'' \cite{Sch49}, p.~54, where $V$ stands for the potential defining the Schr\"odinger equation. Then, considering the harmonic oscillator, he realized that arbitrary superpositions of the related solutions are periodic functions of $t$, with the period of the classical oscillator $\tau_{osc} =2\pi/\omega$. His conclusion on the matter is very interesting to our review purposes: ``this suggest that it might be possible to find a solution in the form of a wave packet whose center of gravity oscillates with the period of the classical motion'' \cite{Sch49}, p.~67. After some calculations, Schiff finally proved that such time-dependent wave packet is viable actually. However, the oscillator wave packet of Schiff was reported by Schr\"odinger 23 years earlier \cite{Sch26c}, as a minor result of his wave--formulation of quantum mechanics \cite{Sch03}, in a try to give a physical meaning to the function that bears his name (see Section~\ref{portrait}). It seems that Schiff was not aware of the Schr\"odinger results at the time of the first edition of his book \cite{Sch49} since the paper \cite{Sch26c} is not mentioned until the third edition \cite{Sch68}, published in 1968. It is also to be noted that the minimum wave packet (\ref{packet}) is as well discussed in the Saxon's book \cite{Sax68}, where the celebrated Schr\"odinger's solution is not mentioned either. According to Saxon, supposing $x=f(t)$ is an integral of the classical equations of motion, it is ``tempting to guess that a suitable form of the corresponding quantum mechanical probability function in the classical limit is $\psi_{\alpha}^* \psi_{\alpha}$ --with $\langle \hat q \rangle$ substituted by $f(t)$ and $\langle \hat p \rangle =0$ in our notation-- for sufficiently small $\Delta \hat q$\,''. He added, ``this expression represents a wave packet of width $\Delta \hat x_1$ moving along the classical trajectory in accordance with the classical equations of motion'' \cite{Sax68}, pp.~26-27. Saxon also proved that (\ref{packet}) is {\em of minimum uncertainty} \cite{Sax68}, pp.~109-110, and then studied the motion of such wave packet in the harmonic oscillator potential \cite{Sax68}, Ch.~VI.6. The analysis by Saxon is very close to that of Schwinger \cite{Sch53}. The above discussion is addressed to emphasize that in late sixties the currently famous paper of Schr\"odinger \cite{Sch26c} was not central in quantum theory yet. Even more, neither the Brown--Twiss experimental results nor the Glauber theory had impacted with enough strength in the literature.

%----------------------------------------------------------->
\subsection{Schr\"odinger's wave packets of minimum uncertainty}
\label{swp}

To recover the Schr\"odinger's wave packet \cite{Sch26c} let us reproduce the steps that led us to Eq.~(\ref{wave3}), but this time using the time-dependent KGS-state $\vert \alpha(t) \rangle$ as point of departure. With the help of  (\ref{alphat}), it is easy to verify the following result
\be
\psi_{\alpha}(x, t) = \frac{e^{-i \omega t }}{\pi^{1/4}} \exp \left\{ - \frac{ [x - \lambda_1(t) ]^2}{2} + i \lambda_2(t) [x - \lambda_1(t)]  t\right\},
\label{wave4}
\ee
where $\lambda_1(t)$ and $\lambda_2(t)$ are the real and imaginary parts of $\alpha \exp(-i\omega t)$, written in short notation as
\be
\vec \lambda(t) =  \frac{1}{\sqrt 2} R(t)  \langle \hat{\vec  x} \rangle, \quad R(t) = \left(
\begin{array}{cc}
\cos \omega t & \sin \omega t \\
-\sin \omega t & \cos \omega t
\end{array}
\right), \quad \vec A = \left(
\begin{array}{c}
A_1\\A_2
\end{array}
\right).
\label{rota}
\ee
The rotation matrix $R(t)$ has the classical oscillator period $\tau_{osc} = 2\pi/\omega$, so the point $x = \lambda_1 (t)$ describes a circumference of radius $\Delta \hat x_1 = \Delta \hat x_2 = 1/\sqrt 2$, centered on the origin of the quadrature phase-space.

Both, Schr\"odinger and Schiff considered a wave packet (\ref{wave4}) with $\langle \hat x_2 \rangle=0$ (the expected value of the initial magnetic-field quadrature is equal to zero!). Notedly, Schr\"odinger was only interested on the real part of his wave packet $\psi_{\alpha}(x,t)$. After dropping the imaginary part of (\ref{wave4}) he wrote (in our notation):
\be
\begin{array}{c}
\psi_{\alpha} (x,t)= \frac{e^{-\frac12 \left( x - \frac{ \langle \hat x_1 \rangle}{\sqrt 2} \cos \omega t \right)^2 } }{\pi^{1/4}}
\cos \left[ \omega t + \left( \frac{ \langle  \hat x_1 \rangle}{\sqrt 2} \sin \omega t \right) \left( x -  \frac{ \langle  \hat x_1 \rangle}{\sqrt 2} \cos \omega t
\right) \right].
\end{array}
\label{schro}
\ee
Schr\"odinger realized that the first factor of (\ref{schro}) represents ``a relatively tall and narrow {\em hump}, of the form of a {\em Gaussian error-curve}, which at a given moment lies in the neighbourhood of the position''  $x= \frac{ \langle  \hat x_1 \rangle}{\sqrt 2} \cos \omega t$ \cite{Sch03}, p.~43. Accordingly, he insisted, ``the hump oscillates under exactly the same law as would operate in the usual mechanics for a particle having (\ref{osc}) as its energy function''. 

The reason for which Schr\"odinger discarded the imaginary part of $\psi_{\alpha}(x,t)$ is that, initially,  he considered the solutions of his equation to be real. Indeed, the factor $i = \sqrt{-1}$ was missing in the first three papers of his celebrated series on {\em quantization as a problem of proper values} \cite{Sch03}. The purely imaginary number was introduced in his fourth paper on quantization. The latter influenced Born to formulate the appropriate probability interpretation of $\psi^* \psi$ \cite{Bor26a}. In his first intuitive intent, Born postulated the probabilities proportional to the probability--amplitudes, just because the papers published by Schr\"odinger at the time formulated $\psi$ to be real. An improved, though still imprecise version (probabilities = amplitude squares), was included at the last moment, in his note added in proof \cite{Mie09}. The argument for such a correction was that (real) amplitudes may be either positive or negative, and the latter cannot be associated with probabilities. Once Born was aware of the complex-valued wave functions, introduced in the Schr\"odinger's fourth paper on quantization, the correct formula was finally provided in \cite{Bor26b}, with some additional precisions by Pauli \cite{Pau27}.

Coming back to the wave packet $\psi_{\alpha}(x,t)$, Schiff based his analysis on the probability density:
\be 
\vert \psi_{\alpha}(x, t) \vert^2 = \pi^{-1/2} \exp \left\{ - \frac{ \left[ \sqrt 2 x - \langle \hat x_1 \rangle \cos \omega t - \langle \hat x_2 \rangle \sin \omega t \right]^2}{2}
\right\},
\label{Schiff}
\ee
which in our case corresponds to the probability of finding the field state at a given eigenvalue $x$ of the (electric-field) quadrature $\hat x_1$. Clearly, the wave packet $\psi_{\alpha}(x, t)$ oscillates without change of shape about $x=0$, with (classical) frequency $\nu_{osc}=\omega/(2\pi)$. 

%----------------------------------------------------------->
\subsection{Time-dependent oscillator wave packets}
\label{twp}

The study of `minimum wave packets' includes a big number of relevant works throughout different stages of modern quantum theory. A non exhaustive list comprehends the pioneering paper of Schr\"odinger \cite{Sch26c}, followed almost immediately by Kennard \cite{Ken27}, where a wave packet is constructed to follow the classical motion with a Gaussian profile but the width of which oscillates with time (the Kennard's state, obtained in 1927, is in this form the first antecedent of what is nowadays called squeezed state!). A very useful ansatz to construct Gaussian wave packets characterized by the exponentiation of quadratic forms was given in 1953 by Husimi \cite{Hus53a,Hus53b}. It was also found that displaced versions of the number states are able to follow the classical motion by keeping their shape \cite{Sen54,Ple54,Ple55,Inf55,Ple56,Eps59} but, unlike the KGS-states, they produce uncertainty products that ``can be arbitrarily large, showing that the classical motion is not necessarily linked with minimum uncertainty'' \cite{Roy82} (see also \cite{Nie97}). A very interesting class of minimum wave packets was exhaustively studied by Nieto \cite{Nie79a,Nie79b,Nie79c,Nie80,Gut80,Nie81}. On the other hand, the dynamics of many physical systems can be described by using the quantum time-dependent harmonic oscillator \cite{Cum86,Cum88,Ghe92,Mih09,Maj05,Leo16,Zha16,Con16,Zel17,Con17,HCr18}, where the construction of minimum wave packets is relevant \cite{Har82,Con11,Com12,Afs16,Mih18,Una18,Zel18,Raz18a}. In a more general situation, wave packets with time-dependent width may occur for systems with different initial conditions, time-dependent frequency, or in contact with a dissipative environment \cite{Cas13a,Cas13b,Cas15,Cas16}. In all these cases, the corresponding coherent states, position and momentum uncertainties, as well as the quantum mechanical energy contributions, can be obtained in the same form if the creation and annihilation operators are expressed in terms of a complex variable that fulfills a nonlinear Riccati equation, which determines the time-evolution of the wave packet width. Explicitly, the wave packet (\ref{wave4}) may be generalized to the form
\be
\Psi (x,t)= N(t) \exp \left\{ i \left[ y(t) \widetilde x^2 + \langle \hat x_2 \rangle \widetilde x + K(t)
\right] \right\},
\label{mipacket}
\ee
where $y(t) = y_R(t) + i y_I(t)$ is a time-dependent complex-valued function, $\widetilde x = x- \langle \hat x_1 \rangle (t) = x - \eta (t)$, with $\eta(t)$ the (dimensionless) position of the wave-packet maximum describing a classical trajectory determined by the Newtonian equation
\be
\ddot \eta(t) + \omega^2 (t) \eta(t) =0,
\label{Newton}
\ee
and $\langle \hat x_2 \rangle = \dot \eta(t)$ the (dimensionless) classical momentum \cite{Cas13a,Cas13b,Cas15,Cas16}. The time-dependent coefficient of the quadratic term obeys the Riccati equation
\be
\dot y(t) + y^2(t) + \omega^2(t)=0.
\label{Riccati}
\ee
The concrete form of the normalization factor $N(t)$ and the purely time-dependent phase $K(t)$ are determined in each case. The straightforward calculation shows that (\ref{Riccati}) is solved by the function $y = \frac{\dot \alpha}{\alpha} + \frac{i}{\alpha}$, where $\alpha$ is a solution of the Ermakov equation
\be
\ddot \alpha(t) + \omega^2 (t) \alpha(t) = \frac{1}{\alpha^3 (t)}.
\ee
The solutions of the above equations depend on the physical system under consideration and on the (complex) initial conditions. Besides, they have close formal similarities with general superpotentials leading to isospectral potentials in supersymmetric quantum mechanics \cite{Coo95,Bag01,Aoy01a,Aoy01b,Mie04,And12,Gan18}. Recent applications include propagation of optical beams in parabolic media \cite{Cru17,Gre17,Raz18b} and Kerr media \cite{Rom15,Rom17,Rom16,Leo15} as well, studies of the geometry of the Riccati equation \cite{Luc16} and the fourth-order Schr\"odinger equation with the energy spectrum of the P\"oschl-Teller system \cite{Reg17}. Further discussion on the subject can be found in the Schuch's book on a nonlinear perspective to quantum theory \cite{Sch18}.

%----------------------------------------------------------->
\subsection{A quantum-family portrait}
\label{portrait}

The development of modern quantum mechanics started in 1925, when Heisenberg conceived the idea that rather constructing a theory from quantities which could not be observed (like the electron orbits inside the atom), one should try to use quantities that are provided by experiment (like the frequencies and amplitudes associated with emission radiation) \cite{Hei71}. Shortly afterwards, during 1925-1927, great progress was made in developing matrix--mechanics  (Heisenberg, Born, Jordan), wave--mechanics (Schr\"odinger), and quantum--algebra (Dirac). The Born's probability interpretation \cite{Bor26a,Bor26b,Pau27} and the uncertainty principle of Heisenberg \cite{Hei27} completed the foundations of what would become the most successful branch of physics throughout the past century. Notwithstanding, interpretative problems arose as soon as the first of the above formulations came to light.  Schr\"odinger, for example, was ``discouraged'' with respect to the ``very difficult methods of transcendental algebra'' of the Heisenberg's theory \cite{Sch26c}. Like Einstein and others, he did not feel comfortable with the quantum jumps involved in the matrix picture. Inspired by the de~Broglie formulation of matter waves (1924), Schr\"odinger introduced a formulation based on finite  single-valued functions obeying an `enigmatic' time-dependent wave equation that is characterized by linear (instead of quadratic) variations in the time-variable \cite{Sch03}. Remarkably, Schr\"odinger was firmly convinced that the solutions of his equation should represent something physically real. In the best of his attempts to provide the wave function with a physical meaning, for the charge density of an electron as a function of the space and time variables, Schr\"odinger suggested the squared modulus of the wave function multiplied by the total charge $e$ (see paper IV on quantization in \cite{Sch03}). From the Schr\"odinger's perspective, the substitution of discrete energies by wave eigenfrequencies would eradicate `quantum jumps' from physics forever. Certainly, he was not immediately aware that not only the Heisenberg's theory and his own formalism are just two faces of the same coin, but that the discreteness of quantum energies as well as quantum jumps arrived to stay. The former was discovered by Schr\"odinger himself (see third paper in \cite{Sch03}), and the later is a natural consequence of the equivalence between both approaches. Ironically, very far from having solved the `problems' of the Heisenberg's picture, the Schr\"odinger's formulation added a number of elements to the quantum theory that are unclear even today, such as the wave function. A main example of the latter can be found in a paper by Dirac \cite{Dir65}, published in 1965. After a detailed comparison between the advantages and disadvantages of the Heisenberg and Schr\"odinger pictures to set up quantum electrodynamics, Dirac found that the two pictures are not equivalent. His opinion is rather clear ``we now see that, if we want a logical quantum electrodynamics, we must work entirely with $q$ numbers in the Heisenberg picture. All references to Schr\"odinger wave functions must be cut out as dead wood. The Schr\"odinger wave functions involve infinities, associated with $v$-$v$ Feynman diagrams, which destroy all hope of logic''. Nevertheless, the above expression requires some caution, as Dirac indicates ``of course the development of quantum theory proposed here should not be considered as detracting from the value of Schr\"odinger's work...  Only when one goes to an infinite number of degrees of freedom does one find that the Schr\"odinger picture is inadequate and that the Heisenberg picture has more fundamental validity.''

%----------------------------------------------------------->
\subsection{The quantum-pictures controversy: hydrogen atom wave packets}
\label{controversy}

The controversy on the Schr\"odinger's quantization papers was immediate. After receiving copies of the first three papers, Lorentz wrote to Schr\"odinger on May 27, 1926, expressing ``if I had to choose now between your wave mechanics and the matrix mechanics, I would give the preference to the former, because of its greater intuitive clarity, so long as one only has to deal with the three coordinates $x,y,z$. If, however, there are more degrees of freedom, then I cannot interpret the waves and vibrations physically, and I must therefore decide in favor of matrix mechanics'' \cite{Prz63}. Clearly, Lorentz was foreseeing the Dirac's arguments on the infinities derived from the Schr\"odinger picture. In addition, among other points, Lorentz remarked that the equation proposed by Schr\"odinger in his first three papers did not contain time-derivatives (namely, it was stationary only), and expressed some doubts about the Schr\"odinger's functions ``if I have understood you correctly, then a {\em particle}, an electron for example, would be comparable to a wave packet which moves with the group velocity. But a wave packet can never stay together and remain confined to a small volume in the long run. The slightest dispersion in the medium will pull it apart in the direction of propagation, and even without that dispersion it will always spread more and more in the transverse direction. Because of this unavoidable blurring a wave packet does not seem to me to be very suitable for representing things to which we want to ascribe a rather permanent individual existence'' \cite{Prz63}. Lorentz also emphasized some difficulties arising in the case of the hydrogen atom and included some calculations on the matter.

On June 6, Schr\"odinger replied that he had found a form to include time-derivatives in his equation. Concerning the difficulties about his functions, he added ``allow me to send you, in an enclosure, a copy of a short note in which something is carried through for the simple case of the oscillator which is also an urgent requirement for all more complicated cases, where however it encounters great computational difficulties. (It would be nicest if it could be carried through in general, but for the present that is hopeless.) It is a question of really establishing the wave groups (or wave packets) which mediate the transition to macroscopic mechanics when one goes to large quantum numbers. You see from the text of the note, which was written before I received your letter, how much I too was concerned about the “staying together” of these wave packets. I am very fortunate that now I can at least point to a simple example where, contrary to all reasonable conjectures, it still proves right'' \cite{Prz63}. 

By `the short note' Schr\"odinger meant his wave packet paper \cite{Sch26c} which, at the very end, included a controversial statement ``we can definitely foresee that, in a similar way, wave groups can be constructed which move around highly quantised Kepler ellipses and are the representation by wave mechanics of the hydrogen electron. But the technical difficulties in the calculation are greater than in the especially simple case which we have treated here'' \cite{Sch03}, p.~44. The latter attracted Lorentz' attention who, on 19 June, replied ``you gave me a great deal of pleasure by sending me your note, {\em the continuous transition from micro- to macro-mechanics}, and as soon as I had read it my first thought was: one must be on the right track with a theory that can refute an objection in such a surprising and beautiful way. Unfortunately my joy immediately dimmed again; namely, I cannot comprehend how, e.g. in the case of the hydrogen atom, you can construct wave packets that move like the electron (I am now thinking of the very high Bohr orbits). The short waves required for doing this are not at your disposal''. Then Lorentz recalled that some lines were wrote by him in his previous communication and proceeded to extend his arguments on the subject. The {\em short note}  of Schr\"odinger \cite{Sch26c} was published on July, 1926. 

The first published criticism appeared in a paper by Heisenberg \cite{Hei27}, received on March 23, 1927. Yes, it is the work in which Heisenberg introduced the uncertainty principle of quantum mechanics! In his own words ``the transition from micro to macro mechanics has already been dealt with by Schr\"odinger \cite{Sch26c}, but I do not believe that Schr\"odinger's considerrations address the essence of the problem'' \cite{Hei27}, p.~184. Heisenberg based his argument on the fact that, unlike the harmonic oscillator, ``the frequencies of the spectral lines emitted by the atom are never integer multiples of a fundamental frequency, according with quantum mechanics with the exception of the special case of the harmonic oscillator. Thus, Schr\"odinger's consideration is applicable only to the harmonic oscillator considered by him, while other cases in the course of time the wave packet spreads over all space surrounding the atom'' \cite{Hei27}, p.~185. That is, Heisenberg criticism was in complete agreement with the doubts expressed by Lorentz. According to Moore \cite{Moo89}, p.~216, Schr\"odinger soon de-emphasized the wave packet picture while Lorentz thought that the demise of wave packets also meant the end of the analogy between wave mechanics and wave optics. 

Taking into account the historical development, we have to say that the work of Schr\"odinger on the oscillator wave packets withstood the test of time just because the Glauber theory came to light. Although Schr\"odinger foresaw some possibilities for his wave packets to be applied in optics, his main efforts were addressed not to solve a practical problem originated in optics, but to provide the solutions of his equation with a physical meaning. The fantastic coincidence of the Schr\"odinger's wave packet and the $x_1$-quadrature representation of the fully coherent states of Glauber is due to the fact that, in both cases, connection with the classical world was looked for the quantum states of the harmonic oscillator. However, the states of Schr\"odinger and those of Glauber were originated by different reasons and obeying different approaches. In this form, the voices declaring that Schr\"odinger `discovered' the coherent states mislead the physical meaning of quantum optical coherence. Simply, it was not possible for Schr\"odinger to guess in any way that his wave packets would be connected with interference phenomena, not even with the Young's interferometer, since no experimental evidence of higher-order coherence (like the Brown--Twiss effect) was available at the time. The connection between the Schr\"odinger's wave packet and the ``position'' representation of the KGS-states is merely mathematical in origin. Nevertheless, the above does not discard the brilliant intuition of Schr\"odinger as far as quantum  physics is concerned.

The works quoted in Section~\ref{owp} testify the correctness of the Lorentz--Heisenberg argument. Although almost all of them are based on the oscillator number states, only the Schr\"odinger wave packets (i.e., the KSG-states) satisfy the properties of preserving their initial Gaussian profile by following the classical oscillator's trajectory. The most `exotic' of the aforementioned oscillator-like systems is the forced oscillator discussed by Husimi \cite{Hus53a,Hus53b}, which was recovered by Carruthers and Nieto as the first application of the KGS-states to study quantum systems other than the harmonic oscillator \cite{Car65}. Nevertheless, attracted by the Schr\"odinger's affirmation, some authors have presented different proposals to solve the problem of finding non-deformable wave packets for the hydrogen atom (see, e.g. \cite{Bro73,Mot77,Bha86,Gay89,Nau89,Gae90,Yea91,Zal94}). Current interest on the subject is addressed to the potential applications of the Rydberg atoms in micro-wave cavities \cite{Har06}. Notably, most of such proposals are based on the dynamics (in the sense of Lie theory) associated to the Coulomb problem. However, as Heisenberg indicated, the states so constructed do not go into a state of the same class under time evolution. The problem was partially solved by Klauder in 1996 \cite{Kla96}, after relaxing some of the properties associated with the KGS-states to construct the appropriate generalized coherent states (see also \cite{Maj97}).

%----------------------------------------------------------->
\section{Group approach and squeezed states }
\label{groupw}

The foundations of property (II) rest on the Lie group-representation of the Heisenberg-Weyl algebra. Namely, the operator $D(\alpha)$ that generates the displaced state $\vert \alpha \rangle = D(\alpha) \vert 0 \rangle$ is obtained by the product of exponentiated forms of the algebra generators $\mathbb I$, $a^{\dagger}$, and $a$. For if we take the exponentiation of $\alpha a^{\dagger}$ and $-\alpha^* a$, using the Baker-Campbell-Hausdorff formula \cite{Mie70}, the related product can be simplified as follows
\be
e^{\alpha a^{\dagger} } e^{-\alpha^* a } = e^{\alpha a^{\dagger} - \alpha^* a} e ^{\frac12 [\alpha^* a , \alpha a^{\dagger}]} = e^{\alpha a^{\dagger} - \alpha^* a} e^{\vert \alpha \vert^2 \mathbb I}.
\ee
Therefore
\be
e^{-\vert \alpha \vert^2 \mathbb I} e^{\alpha a^{\dagger} } e^{-\alpha^* a } = e^{\alpha a^{\dagger} - \alpha^* a}.
\label{d3}
\ee
The factors at the left hand side of (\ref{d3}) are the result of a parametric exponentiation of the generators of the Heisenberg-Weyl algebra. In turn, the element at the right hand side is the exponentiated form of the operator $\alpha a^{\dagger} - \alpha^* a$, which is also a member of the algebra. Notably, all the exponentiated forms included in (\ref{d3}) are elements of the Heisenberg-Weyl group, with $\exp( \alpha a^{\dagger} - \alpha^* a)$ the result of multiplying the group basis elements properly parameterized. Comparing with (\ref{d1}) we find the origin of the displacement operator $D(\alpha)$. On the other hand, expression (\ref{d3}) is a {\em disentangling formula} \cite{Gil74a,Ban93,Das96} that permits the factorization of $D(\alpha)$ into a normally ordered form, see Eq.~(\ref{normal}). The factorization in antinormal-order is also feasible \cite{Gil74a,Ban93,Das96}. As the zero-photon state $\vert 0 \rangle$ is annihilated by $a$, it is immediate to see that the action of the group-element $\exp( -\alpha^* a)$ leaves $\vert 0 \rangle$ invariant. Hence $\vert 0 \rangle$ is called {\em fiducial state} with respect to $\exp( -\alpha^* a)$ \cite{Per72,Per77,Per86,Gil74b,Zha90}. The above results can be abridged by saying that the Heisenberg-Weyl algebra rules the dynamics of the quantum harmonic oscillator, and that the basis elements of the related Lie group define a mechanism to get the coherent state $\vert \alpha \rangle$, with the vacuum $\vert 0 \rangle$ as fiducial state. Then we have at hand a recipe to generalize the notion of coherent states for which the symmetric properties of the system under study are relevant. It is just required the identification of the dynamical group that defines the properties of the system  we are interested in, as well as the related fiducial state \cite{Per72,Per77,Per86,Gil74b,Zha90}. 

%----------------------------------------------------------->
\subsection{The versatile representations of $SU(1,1)$ and $SU(2)$ Lie groups}

To give an example let us consider the operators $K_{\pm}$ and $K_0$ that satisfy the commutation relations
\be
[K_-,K_+]= 2\sigma_{\pm} K_0, \quad [K_0, K_{\pm} ]= \pm K_{\pm}, \quad \sigma_{\pm} =\pm 1,
\ee
as well as the operator
\be
K^2 = K_0^2 - \sigma_{\pm}^{1/2} (K_+ K_- + K_-K_+),
\ee
which satisfies $[K^2, K_{\pm}]=[K^2, K_0]=0$. The above expressions correspond to the $su(1,1)$ Lie algebra for $\sigma_+$, and to the $su(2)$ Lie algebra for $\sigma_-$, with $K^2$ the Casimir operator. The normal order disentangling formula is in this case as follows \cite{Ban93}:
\be
e^{A_+ K_+} e^{ (\ln A_0) K_0} e^{A_- K_-} = e^{a_+ K_+ + a_0 K_0 + a_- K_-},
\label{d4}
\ee
with
\[
A_{\pm} = \frac{a_{\pm} \sinh \phi}{\phi \sqrt A_0}, \quad A_0 = \frac{1}{ [ \cosh \phi -  \frac{a_0 \sinh \phi}{2 \phi}]^2 }, \quad \phi = \left[ \frac{a_0^2}{4} - \sigma_{\pm} a_+ a_-
\right]^{1/2}. 
\]
For the $su(1,1)$ Lie algebra we have two immediate bosonic representations. Namely, the single-mode one
\be
K_+ = \frac12 (a^{\dagger})^2, \quad K_-= \frac12 a^2, \quad K_0= \frac12 \left(a^{\dagger} a  + \frac12 \right),
\label{1mode}
\ee
and the two-mode representation
\be
K_+ = a^{\dagger} b^{\dagger}, \quad K_- = ab, \quad K_0 =\frac12 \left( a^{\dagger} a + b^{\dagger} b + 1\right),
\label{2mode}
\ee
where $[a, a^{\dagger}]=[b, b^{\dagger}]=1$, $[a, b^{\dagger} ]=0$, etc. 

%----------------------------------------------------------->
\subsubsection{Vacuum squeezed states}

From (\ref{d4}) and (\ref{1mode}), with $a_0=0$, $a_+ = \xi = - a_-^*$, and $\xi = r e^{i\varphi} \in \mathbb C$, we obtain a new operator $S_s(\xi) = \exp [\xi (a^{\dagger})^2 - \xi^* a^2 ]/2$ \cite{Sto70a,Sto71} such that 
\be
\vert \xi_s \rangle := S_s(\xi) \vert 0 \rangle = \frac{1}{\sqrt{ \cosh r} } \sum_{n=0}^{\infty} \frac{ e^{in\varphi} \tanh^n r }{2^n n!} \sqrt{(2 n)!} \vert 2 n \rangle.
\label{squeezed}
\ee
The superposition $\vert \xi_s \rangle$ is known as the vacuum squeezed state \cite{Hol79} (see also \cite{Wal83,Lou87,Tei89}). The first antecedents of these states can be found in the paper of Kennard \cite{Ken27}, and those of Infeld and Pleb\'anski \cite{Ple54,Ple55,Inf55,Ple56}. Noticeably, Cachill and Glauber used a mixed state prepared with even number states $\vert 2n \rangle \langle 2n \vert$ to explore a class of displaced thermal states \cite{Gla07}, Ch.~10. The pure state $\vert \xi_s \rangle \langle \xi_s \vert$ includes populations $\vert 2n \rangle \langle 2n \vert$ as well as coherences $\vert 2n \rangle \langle 2m \vert$, $n \neq m$, and the former may share some properties with the Cachill-Glauber states since it corresponds to the `classical' part of the vacuum squeezed state (\ref{squeezed}).

On the other hand, it may be proved that the Mandel parameter (\ref{mandel}) is in this case $Q_M= 2 \langle \hat n \rangle +1 = 2 \sinh^2 r + 1$, and that the variances (evaluated with arbitrary $\varphi$) are parameterized by the modulus of $\xi$  as follows $(\Delta \hat x_1 )^2 = e^{2r}/2$, $(\Delta \hat x_2 )^2 = e^{-2r}/2$. Clearly, although $\hat x_1$ is stretched at the time that $\hat x_2$ is squeezed, the related uncertainty is minimized. The roles are interchanged for other values of $\varphi$. From (\ref{corr4}) we also see that $g^{(2)} = 3 + 1/\sinh^2r$. Therefore, as $g^{(2)} >3$, the vacuum squeezed state $\vert \xi_s \rangle$ is nonclassical. The latter is enforced by noticing that $Q_M >1$. Of course, $r=0$ is forbidden to evaluate either $g^{(2)}$ or $Q_M$ since this produces the vacuum state $\vert \xi_s = 0 \rangle$. Additionally, the state $\vert \xi_s \rangle$ is temporally stable $\vert \xi_s (t) \rangle = e^{-i \omega t/2} \vert \xi e^{-2 i\omega t} \rangle$. On the other hand, the expectation value $\langle a^2 \rangle = \sinh (2r) e^{i\varphi} /2$ leads to the probability $\vert \langle a^2 \rangle \vert^2 = \sinh^2 (2r)/4$ that two photons have been detected at the same space point by preserving the field state. This probability is always different from zero and increases with the mean number of photons.

%----------------------------------------------------------->
\subsubsection{Even and odd coherent states}

Remarkably, Dodonov, Malkin and Man'ko introduced in 1974 a pair of states $\vert \alpha_{\pm} \rangle$, called {\em even and odd coherent states} \cite{Dod74}, which include (\ref{squeezed}) as the even case $\vert \alpha_+\rangle$. These states result from the superpositions 
\be
\vert \alpha_{\pm} \rangle = \left[ 2 \left(1 \pm e^{-2 \vert \alpha \vert^2} \right) \right]^{-1/2} (\vert \alpha \rangle \pm \vert - \alpha \rangle).
\ee
The properties of $\vert \alpha_{\pm} \rangle$ can be consulted in the review by Man'ko \cite{Dod03}, Ch.~4. For historical details about general squeezed states see the pretty review prepared by Dodonov and Man'ko \cite{Dod03}, Ch.~1. 

%----------------------------------------------------------->
\subsubsection{Squeezed coherent states}

Applying the displacement operator $D(\alpha)$ on $\vert \xi_s \rangle$ we obtain the squeezed coherent state $\vert \alpha, \xi_s \rangle = D(\alpha) S_s(\xi) \vert 0 \rangle$ \cite{Yue76,Lou87}, which produces the same variances as $\vert \xi_s \rangle$, with a modification in the expected number of photons $\langle \hat n \rangle = \vert \alpha \vert^2 + \sinh^2 r$. Thus, the displacement produced by $D(\alpha)$ does not modify the squeezing properties. However, depending on the combination of $\alpha$ and $r$, the Mandel parameter $Q_M$ can be negative and $g^{(2)} \leq 1$ as well. In other words, for such values of $(r, \alpha)$ the states $\vert \alpha, \xi_s \rangle$ are very close in properties to the number states $\vert n \geq 1\rangle$.

%----------------------------------------------------------->
\subsubsection{Two-mode squeezed states}

If we now use (\ref{d4}) and (\ref{2mode}) with the same parameters $a_0$, $a_{\pm}$ as before, the action of the resulting operator $S_{two}(\xi) = \exp (\xi a^{\dagger} b^{\dagger} - \xi^* a b )$ on the two-mode vacuum state $\vert 0,0 \rangle$ produces \cite{Cav85,Sch85}:
\be
\vert \xi_{two} \rangle = \frac{1}{\cosh r} \sum_{n=0}^{\infty} e^{i n \varphi} \tanh^n r \vert n, n \rangle.
\ee
The mean number of photons in any mode is the same $\langle \hat n_a \rangle = \langle \hat n_b \rangle =\sinh^2 r$, with correlation between modes defined by $\langle ab \rangle = \sinh (2r) e^{i\varphi}/2$ and $\langle a^{\dagger} b \rangle=0$. Besides $\langle a^2 \rangle = \langle b^2 \rangle =0$, and $\langle a \rangle = \langle b \rangle=0$. The latter means that detection of only one photon, no matter the mode, is forbidden. Indeed, as $\vert \langle a \rangle \vert^2 = \vert \langle b \rangle \vert^2 =0$, we see that the state of the field is inevitably changed to an orthogonal configuration after detecting a single photon! The same holds for the probabilities $\vert \langle a^2 \rangle \vert^2 = \vert \langle b^2 \rangle \vert^2 =0$, so the field is changed to an orthogonal configuration after detecting two photons of the same mode. The situation is different for the square modulus of the correlation $\vert \langle ab \rangle \vert^2 = \sinh^2 (2r)/4$, which means that only detections of one photon in a given mode AND one photon in the complementary mode are allowed. These properties are such that the modes $a$ and $b$ themselves are not squeezed, and suggest the symmetrization $(a \pm b)/{\sqrt 2}$ to obtain the properly defined quadratures (a 50-50 beam splitter is useful in this subject \cite{Aga13}, Ch.~3.2). Then, it may be shown that the squeezing operator $S_{two}(\xi)$ can be factorized as the product of two-single mode squeezing operators. It is also possible to define two-mode squeezed coherent states $\vert \alpha, \beta, \xi \rangle = D(\alpha) D(\beta) S_{two}(\xi ) \vert 0,0 \rangle$. Additional properties of these states can be consulted in the book by Agarwal \cite{Aga13}.

%----------------------------------------------------------->
\section{Generalized coherent states}
\label{gral}

The bare essentials of coherent states can be expressed as a linear superposition 
\be
\vert \alpha_{CS} \rangle = \sum_{k \in {\cal I}} f_n (\alpha) \vert \psi_n \rangle, \quad \alpha \in \mathbb C,
\label{g1}
\ee
where the vectors $\vert \psi_n \rangle$ generate a (separable) Hilbert space ${\cal H}$ \cite{Gla94}, ${\cal I} \subseteq \mathbb R$ is an appropriate set of indices, and $f_n(\alpha)$ is a set of analytical functions permitting normalization. The superposition (\ref{g1}) must exhibit some specific properties that are determined by the `user' with basis on either the phenomenology under study or theoretical arguments. 

Although the Klauder-Glauber-Sudarshan states (\ref{glauber1}) are the only superpositions that posses all the properties (A), (B), (I)--(III), and many other, the term {\em coherent states} (CS) has been used for a wide class of mathematical objects over the years. Nowadays, the overcomplete bases of states constructed to include at least one of the properties discussed in Section~\ref{KGS} are called {\em generalized coherent states}. The most valuable profile of the latter is that they can be studied for many systems in terms of the definition leading to the desirable result. For instance, the generalized CS studied by Barut and Girardello \cite{Bar71} and by Perelomov \cite{Per72,Per77,Per86} are respectively based on properties (A) and (II) of the KGS-states. Property (III) is incidentally found as a secondary result for some special systems. Indeed, besides the Schr\"odinger \cite{Sch26c}, Kennard \cite{Ken27}, Schiff \cite{Sch49}, Husimi \cite{Hus53a,Hus53b},  and Saxon \cite{Sax68} contributions, the construction of wave packets  addressed to minimize the uncertainty relation of a pair of observables has been rarely reported in the literature. An exception is represented by the Nieto's results \cite{Nie79a,Nie79b,Nie79c,Nie80,Gut80,Nie81}. Of course, we have in mind that squeezed states can be considered as deformations of generalized CS for which the quadratures are not equal but minimize the related uncertainty. Klauder, for instance, constructed generalized CS for the hydrogen atom \cite{Kla96} such that they have temporal estability (i.e., they satisfy property I), are normalized and parametrized continuously, like it is defined in Eq.~(\ref{g1}), and admit a resolution of unity with a positive measure (a fundamental property of the KGS-states proved in advance by Klauder himself \cite{Kla63a,Kla63b}). Thus, as a basis to construct his states for the hydrogen atom, Klauder used the concept of `coherent state' introduced in his compilation book, signed together with Skagerstam \cite{Kla85}. Further improvements were given in \cite{Gaz99}. Additional generalizations have been discussed in, e.g. \cite{Fon88,Ali95,Ali99}. 

As a general rule, it is expected that the set $\{ \vert \alpha_{CS} \rangle \}$ forms an overcomplete basis of the corresponding Hilbert space ${\cal H}$. In turn, the superpositions $\vert \alpha_{CS} \rangle$ are wished to be temporally stable. 

%----------------------------------------------------------->
\subsection{Generalized oscillator algebras}

The deformed oscillator (or boson) algebras can be encoded in a global symbolic expression that facilitates their study. One of the advantages of working in symbolic form is that the related nonlinear coherent states can be written in the same mathematical context \cite{Zel17a}. The main idea is to consider (\ref{g1}) with
\be
f_n(\alpha) = \frac{\alpha^n}{\sqrt{E(n)!}}, \quad E(n)!= E(1) E(2) \cdots E(n), \quad E(0)!:=1, 
\label{g2}
\ee
where $E$ is a nonnegative function, ${\cal I} = \{ 0, 1, 2, \ldots\}$, and $\vert \psi_n \rangle \equiv \vert n \rangle \, \forall n \in {\cal I}$. The normalization requires
\be
{\cal N}_E(\vert \alpha \vert) = \left[ \sum_{n=0}^{\infty} \frac{\vert \alpha \vert^{2n} }{E(n)!} \right]^{-1/2}
\ee
to be finite, so that not any $E$ and $\vert \alpha \vert$ are allowed. Assuming $E$ and $\alpha$ properly chosen, the normalized vectors $\vert \alpha_{CS} \rangle_N = {\cal N}_E (\alpha) \vert \alpha_{CS} \rangle$ satisfy the closure relation
\be
\mathbb I = \int   \vert \alpha_{CS} \rangle_N \langle \alpha_{CS} \vert d \sigma_E (\alpha); \qquad 
d \sigma_E (\alpha) = \frac{d^2 \alpha}{\pi} \Lambda_E \left( \vert \alpha \vert^2 \right),
\ee
with $\Lambda_E (\alpha)$ an additional function to be determined such that 
\be
\int_0^{\infty} \Lambda_E (x) x^n dx = E(n)!, \quad \alpha = re^{i\theta}, \quad x=r^2.
\label{measure}
\ee
After the change $b \rightarrow m-1$, integral equation (\ref{measure})  coincides with the Mellin transform \cite{Ber00} of $\Lambda_E (x)$. The simplest form to obtain (\ref{g2}) is by a modification of the oscillator algebra that preserves the number operator $\hat n$ but changes the ladder operators, now written $a_E$ and $a_E^{\dagger}$, as the set of generators. That is, one has
\be
[\hat n, a_E ]= -a_E, \quad [\hat n, a_E^{\dagger} ] = a_E^{\dagger},
\ee
so that the product $a_E^{\dagger} a_E$ preserves the number of quanta provided it is equal to the function $E(\hat n)$. Equivalently, $a_E a_E^{\dagger} = E(\hat n +1)$, so that $a_E \vert n \rangle = \sqrt{E(n)} \vert n-1 \rangle$, $a_E^{\dagger} \vert n \rangle = \sqrt{E(n+1)} \vert n+1 \rangle$, and
\[
[a_E, a_E^{\dagger} ]-E(\hat n+1) =E(\hat n).
\]
As the vacuum state $\vert 0 \rangle$ does not contain quanta we shall assume $E(0) =0$ in order to have a bounded from below annihilation operator $a_E \vert 0 \rangle=0$. It is say that any system obeying the new algebra is a generalized oscillator. It may be shown that when the function $E(n)$ is a real polynomial of degree $\ell \geq 1$, the determination of the measure (\ref{measure}) is reduced to a moment problem that is solved by the Meijer G-function \cite{Zel17a}. This property automatically defines the delta distribution as the $P$-representation of the generalized CS $\vert \alpha_{CS} \rangle_N$ so definded. Then, in principle, there must be a classical analogy for them. However, in \cite{Zel17a} it has been shown that they exhibit properties like antibunching and that they lack second-order coherence. That is, although they are $𝑃$-represented by a delta function, they are not fully coherent. Therefore, the systems associated with the generalized oscillator algebras cannot be considered ``classical''  in the context of the quantum theory of optical coherence. Examples include the $f$-oscillators of Man'ko and co-workers \cite{Man97},  $q$-deformed oscillators \cite{Bie89,Mac89,Ros00}, deƒformed photon phenomenology \cite{Sol94}, the $su(1,1)$ oscillators applied to the study of the Jaynes-Cummings model \cite{Jay63} for intensity dependent interactions \cite{Buc81,Suk81} as well as the supersymmetric and nonlinear models \cite{Mat96b,Man97,Ber93,Fer94,Fer95,Ros96,Kum96,Fer07,Fer99,Fer08,Zel17b,Ros18b,Orz88,Jun99,Roy00a,Rok04a,Rok04b,Tav08,Bag08,Twa08,Bag09a,Bag09b,Hon09,Abb09,Tav10,Saf11,Kor13,Noo14,Hus11,Ang12,Ang13,Zel17a,Moj18}.

%----------------------------------------------------------->
\subsection{Position--dependent mass systems and quantum--classical analogies}

The problem of calculating the energies of quantum systems endowed with position-dependent mass $m(x)$ has been a subject of increasing interest in recent years \cite{Bas81,Bas82,Mil99,Sou00,Roy05,Gan06,Mus06,Cru07,Cru08a,Cru08b,Mus09,Cru13,Bag13,Lak13,Cru13b,Cos14,Cru14,Nik15,Mus15,Gos15,Bag15,Ami16a,Scr16,Bra16,Nik17,Cos18,Jes18,Che18,Cru18
}. This model represents an interface between theoretical and applied physics, with analogies in geometric optics where the position dependent refractive index can be interpreted as a variable mass \cite{Wol04}. Its main characteristic is that the conventional expression for the kinetic term $\frac{\hat p^2}{2m}$ is not self-adjoint \cite{Bas81,Bas82}, so that the Hermiticity of the Hamiltonian is a part of the problem if the mass is not a constant. Nevertheless, different generalized CS have been constructed \cite{Cru11,Sou08,Cru09,Rub10,Yah12,Yah14,Ami16b,Yah17,Ami17}. The above is remarkable since well known quantum--classical analogies \cite{Tho94,Sau17,Wol04,Dra04,Rau15,Bel19} can be exploited to test quantum-theoretical predictions in the laboratory. That is, although we nowadays have at hand precise forms to produce single photons on demand (see, e.g.~\cite{Pro15,Cal16,Lop17} and the review paper \cite{Cou18}), relevant information is accessible from the optical analogies of quantum behavior \cite{Dra04,Bel19}. A first example is offered by the propagation of signals in optical waveguides \cite{Oka11}, which can be used to test important predictions dealing with quantum resonances and leaky electromagnetic modes \cite{Cru15R,Cru15b}, solitons \cite{Hon11b,Ros18c} and supersymmetry \cite{Wal18,Con18}. Another example arises by recalling that classic optics includes an uncertainty relation between position and momentum, with Plack's constant $\hbar$ replaced by the light wave-length $\lambda$ \cite{Dra04}. The latter analogies connect, as we have seen, Gaussian light beams with fully coherent states. However, it may be also useful in the study of multilevel quantum systems \cite{Gil75}. Of course, the studies on the propagation of optical beams in parabolic \cite{Cru17,Gre17,Raz18b} and Kerr \cite{Rom15,Rom17,Rom16,Leo15} media include a refractive index with special properties that can be expressed as a concrete parametrization of the quantum states of light. On the other hand, quantum field theory in curved spacetime also permits classical analogies \cite{Bel19}, so that Hawking radiation can be studied in nonlinear Kerr media (including the analysis of the vacuum state for a star collapsing to a black hole which leads to the controversial effect named after Unruh \cite{Cru16a,Coz17,Cet17,Ros18}). At the moment, the analogies of Hawking radiation seem to be feasible \cite{Ber16,Ber18a,Ber18b}.

%----------------------------------------------------------->
\subsection{Two faces of the same coin}

Considering the above remarks, it must be clear that squeezed and coherent states can be treated as different faces of a superposition $\vert \alpha_{CS} \rangle$ that minimizes the uncertainty of the appropriate quadratures. Indeed, they appear at different times when repeated measurements are made on a system \cite{Boc96}, in the context of quantum nondemolition measurements \cite{Bra80,Cav80a,Cav80b} required to detect gravitational waves \cite{Tho94,Sch10,Sau17}, where the expression {\em squeezed state} was coined \cite{Hol79} (see also \cite{Cle10,Gue07,Say11}). Applications include quantum gravity \cite{Ori12}, gauge field theory \cite{Thi01a,Thi01b,Thi01c,Thi01d}, and Yang-Mills theory \cite{Hal01}. Addressed to the interest on Rydberg atoms \cite{Har06,Leo78}, the generalized CS \cite{Lee97,Mes85,Bru87,Gal88} are also relevant in manipulating atom-photon interactions \cite{Har06,Win98,Dav99,Rai01,Moy06}. Special properties are revealed in the Jaynes-Cummings model \cite{Ber94,Dao02,Hus05,Fin08}, and for dynamical systems ruled by either the $su(1,1)$ or the $su(2)$ Lie algebras \cite{Cru11,Are72,Gaz92,Alv02,Wun03,Mir12,Kar14,Dan15,Ros16}. The coherent states can be entangled \cite{San12,Li13,Dey15,Mot18}, superposed \cite{Dod74,Ger93,Mat96a,Roy98,Spi95}, and constructed for non-Hermitian operators \cite{Zel17b,Ros18b,Dey18} in terms of either a bi-orthogonal basis \cite{Jai17,Zel17b,Ros18b,Ali04} or noncommutative spaces \cite{Dey15,Dey16,Dey17}, for which nonclassical properties can be found \cite{Zel17b,Her16,Zel17a,Zel18b}. They have been also associated to super algebraic structures \cite{Mot18,Ara86,Gra96,Jay99,Alv04,Nie06}, nonlinear oscillators \cite{Kin01,Mid09,Chi13}, and solvable models \cite{Roy00,Wun06,Tav06,Del07,Ang07,Hon11a,Tav13,Hol14,Hof18,Hof18b,Ant01,Ber11,Gaz12}.

%----------------------------------------------------------->
\section{Conclusion}
\label{concluye}

I have revisited the profiles of coherent and squeezed states that, in my opinion, have been overpassed in other works of the same nature and subject. The efforts have been addressed to clarify the main concepts and notions, including some passages of the history of science, with the aim of facilitating the subject for nonspecialists. In this sense, the present work must be treated as complementary to the reviews already reported by other authors. Clearly, it is not possible to scan all the literature on the matter so that, by necessity, any review is an incomplete work. The papers included in the references cannot cover all the relevant contributions on the matter,  so  the bibliography is, after all, the imperfect selection of the author. I apologize for the missed fundamental references as well as for the imprecise quotations (if any). I would conclude the work by addressing the readers attention to the books by Mandel and Wolf \cite{Man95},  Klauder and Sudarshan \cite{Kla68}, Glauber \cite{Gla07}, Perelomov \cite{Per86}, Ali, Antoine, and Gazeau \cite{Ali99}, Dodonov and Man'ko \cite{Dod03}, Combescure and Robert \cite{Com12}, Agarwal \cite{Aga13}, Schuch \cite{Sch18}, and Gazeau \cite{Gaz09}. The reviews of  Gilmore \cite{Gil74b}, Walls \cite{Wal83}, Loudon and Knight \cite{Lou87}, Fonda, Mankoc-Brostnik and Rosina \cite{Fon88}, Zhang, Feng and Gilmore \cite{Zha90}, and Ali, Antoine, Gazeau and Mueller \cite{Ali95} are also terrific to initiate the study of the subject. 

%---------------------------------------> Section
\section*{Acknowledgment}

Financial support form Ministerio de Econom\'ia y Competitividad (Spain) grant number MTM2014-57129-C2-1-P, Consejer\'ia de Educaci\'on, Junta de Castilla y Le\'on (Spain) grant number VA057U16, and Consejo Nacional de Ciencia y Tecnolog\'ia (Mexico)  project A1-S-24569 is acknowledged.

%--------------------------------------->

\end{document}